\documentclass[11pt,a4paper]{article}
   \usepackage{times}
    \usepackage[english]{babel}
    \usepackage{amsmath,amssymb,amsfonts}
   \usepackage{textcomp}
    \usepackage{graphicx}
    \usepackage[usenames]{color}
    \newcommand{\Section}[1]{\section{\hskip -0.75em ~#1}}

    \renewcommand{\thesection}{\arabic{section}}
    \renewcommand{\thesection}{\arabic{section}}
    
    \newcounter{multieqs}
    

\begin{document}
\author{S\'{i}lvio R. Dahmen\\
\normalsize Instituto de F\'{\i}sica, Universidade Federal do Rio Grande do Sul\\
90540 -- 090 Porto Alegre Brazil}
\title{The Mathematics and Physics of Diderot.\\
I. On Pendulums and Air Resistance\\}
\maketitle{}

\begin{abstract}

In this article Denis Diderot's Fifth Memoir of 1748 on the problem of a pendulum damped by air resistance
is discussed.

Diderot wrote the Memoir in order to clarify an assumption
Newton made without further justification in the first pages of the {\it Principia} in connection with an experiment to verify
the Third Law of Motion using colliding pendulums. To explain the differences between experimental and theoretical values of momentum
in the collision experiments he conducted Newton assumed
that the bob was retarded by an air resistance $F_R$ proportional to the velocity $v$. 
By giving Newton's arguments a mathematical scaffolding and recasting his geometrical reasoning in the language of differential
calculus, Diderot provides a step-by-step solution guide to the problem and proposes experiments to settle the question about the appropriate form of $F_R$,
which for Diderot quadratic in $v$, that is $F_R \sim v^2$.

The solution of Diderot is presented in full detail and his results are compared to those obtained from a Lindstedt-Poincar\'e approximation for an oscillator with quadratic damping. It is shown that, up to a prefactor, both coincide. Some results that one can derive from his approach are presented and discussed for the first time. Experimental evidence to support Diderot's or Newton's claims is discussed together with the
limitations of their solutions. Some misprints in the original memoir are pointed out.

%It is also argued that
%Diderot's work on involutes, apparently unrelated to the present
%problem, reveals a close resemblance to the technical problem involved
%in the construction of an isochronous pendulum clock discussed in
%Huygens's {\it Horologium Oscillatorium}, which could have served as
%his inspiration~\cite{huygenswork}. This can be further atested by Diderot's involvement with the use of involutes in the rectification problem,
%also to be found in Huygens's seminal work. 
\end{abstract}
\newpage
\Section{How to read this article}
This article has two main objectives: first, to discuss the historical context and educational aspects of  Diderot's commentaries on Newton's treatment of the
damped oscillator briefly discussed in the opening sections of the {\it Principia} in connection with the third law of motion~\cite{diderot1}.
Second, to provide a detailed mathematical analysis of Diderot's calculations and compare his results with those obtained  using the modern approach
to the problem of a damped oscillator. As such, the paper has a historical part with a detailed reproduction of Diderot's results
and a section where the modern approach to the damped oscillator is presented and experimental results discussed. 

The historian Peter Gay, in his seminal book on the Enlightenment said that {\it `... Diderot was,
with almost equal competence, translator, editor, playwright, psychologist, art critic and theorist, novelist, classical scholar, and
educational and ethical reformer'}~\cite{gay}. To this one should add, if not {\it mathematician} in the strict sense of the word,
at least the epiteth {\it mathematics enthusiast}. He was no professional mathematician, as his fellow {\it philosophe} Jean De La Ronde D'Alembert was, but his
involvement went beyond that of a simple amateur. Given the range of Diderot's interests and his standing among the {\it philosophes} of the Enlightenment,
his delvings into the fields of mathematics and physics call for a more detailed analysis. However, it is
also true that these very achievements can only be fully appreciated if we compare his approach and results with all that we know today.
The combination of these two facets of his work comes at a cost: the long calculations of Diderot's article
and the modern view on the subject cannot be presented in a condensed way. So I have tried to show the two sides of the coin
in such a way as to allow them to be examined independently, if necessary. By doing this 
it was my purpose to spare those readers interested only in the history from reading the modern approach without, at the same time,
compromising the comprehension of Diderot's work as we understand it today. 

For those interested in the historical aspects of Diderot's Memoirs on Mathematics, one may concentrate on Sections \ref{sec:intro},
\ref{sec:DiderotMathematician} and \ref{sec:DiderotMemoir}. His calculations are presented in detail in Section
\ref{sec:DiderotPerspective}. This is the longest section and some results missing in the original are discussed for
the first time. Section \ref{sec:MathematicalPendulum} is where the modern approach to the damped oscillator is discussed, both from
a theoretical as well as from an experimental viewpoint. As mentioned in the preceeding paragraph, this section may be skipped at a first
reading. Some of the results presented there which help us better understand the complexity of the problem Diderot dealt with and his solution
are clearly indicated along the main text.
%\tableofcontents 
%    \renewcommand{\chaptermark}[1]
%      {\markboth{\sf #1}{}}
    \renewcommand{\sectionmark}[1]
 {\markright{\sf\thesection\ #1}}
\Section{Introduction}
\label{sec:intro}
\noindent
One of the last things to come to mind when one thinks of Denis Diderot (1713--1784) is the field of mathematics and physics.
Rightly regarded as one of the most prolific minds of the 18th century, his name
evokes first and foremost the emblematic {\it Encyclop\'edie des Sciences, des Arts e des M\'etiers}, of which he
was the main editor and to which he dedicated
25 years of his life\footnote{Diderot's association with the {\it Encyclop\'edie}'s started in 1747. The first 17 volumes
of the colossal work were published between 1751 and 1765. The eleven extra volumes of plates were finished
by 1772.}. Given the gargantuan range
of his interests -- encyclopedic in the broadest sense of the word -- one can find under his pen
works of philosophical enquiry, historiography, critique of art, novels and translations. Thus it should come as no surprise
that the specialized and non-specialized
literature on Diderot reflect, in variety and extent, the breadth of his intellectual production. However, given the wealth of
mathematical and physical problems Diderot tried his hand at, the same cannot be said of his mathematical treatises. With a few praiseworthy
exceptions~\cite{kk,coolidge,ballstadt} this facet of Diderot -- that of the mathematician -- remains largely untouched.

One possible reason for the lack of interest could be the fact that some of his mathematical writings are rather technical and deal
with very specific problems of physics and applied mathematics: sundials, probability theory, theory of algebraic curves, calculus
of annuities and deciphering machines, to name just a few.
Moreover, his association with other fields of enquiry are so vast that his
mathematical exploits are -- to a great degree still -- regarded by some as the work of a dilletante, a minor diversion
from his more influential works. This couldn't be less true. His involvement with mathematics,
based on Diderot's own account and his writings extend for a period of 28 years, from 1733 to 1761. His later works are full
of comments on mathematics and the value he accorded to it can be judged by the fact that he placed it at the basis of the curriculum
for a university in Russia\footnote{His {\it Plan d'une universit\'e pour le gouvernement de Russie} (1775) was a
personal request of the Russian Empress Catherine II~\cite{wilson}.}.
As Ballstadt convincingly showed in his long treatise on Diderot and the Natural Sciences, if one wants to
fathom Diderot as a natural philosopher in his entirety it is necessary to understand
the role mathematics played in his thought~\cite{ballstadt}.
When compared to mathematicians like Jean de La Ronde D'Alembert (1717--1783), his coeditor at the Encyclopedia until 1759,
one may rightfully call him an amateur -- but if so, one should add that he was an extremely competent one.
He did original research, was technically proficient and used techniques which,
at that time, were at the very front of the research arena. Thus Diderot the pantophile should not overshadow Diderot the mathematician.

The first to consider mathematics in the context of Diderot's works were Krakeur and Krueger~\cite{kk},
whom we owe much of what we know about the subject today. The first technical analysis of Diderot work in mathematics was undertaken by
Coolidge~\cite{coolidge}. In spite of its shortcomings -- only some of Diderot's works are analysed and then only
partially -- it remains a starting point and a valuable source of information for any serious study. Ballstadt
considers the role mathematics played in Diderot's views on natural philosophy and, given his
encyclopedic bent of mind, which encompassed basically every branch of
knowledge at that time, mathematics can be rightfully said to have been one of his greatest and never fading passions~\cite{ballstadt}. 
Diderot himself referred in his later years to his relation to mathematics in a rather amusing way. In his {\it R\'efutation d'Helv\'etius} (1774) he
tells the story of two parents who, upon noticing that their first-born son was rather predisposed to studies, decided to send him to the local
provincial school and later to Paris, to the University, where:
\begin{quote}
{\it They gave him texts on arithmetics, algebra and geometry, which he devoured. Later, admonished to [devote himself] to
more agreeable studies, he found pleasure in the reading of Homer, Virgil, Tasso and Milton, but always returned to mathematics,
just like an unfaithful husband who, tired of his mistress, returns from time to time to his wife
\footnote{On lui met entre les mains des cahiers d'arithm\'etique,
d'alg\`ebre et de g\'eom\'etrie qu'il d\'evora. Entra\^{\i}n\'e par la suite \`a des \'etudes plus agr\'eables,
il se plut \`a la lecture d'Hom\`ere, de Virgile, du Tasse e de Milton, mais revenant toujours aux math\'ematiques,
comme un \'epoux infid\`ele, las de sa ma\^itresse, revient de temps en temps \`a sa femme~\cite{diderothelvetius}.}.}
\end{quote}
Except for the part where the parents sent the son to Paris, this is his own story\footnote{Diderot actually tried to leave Langres,
his hometown, without the consent of his father to join the Jesuits in Paris. His father got wind of it and took his son himself to Paris,
enrolling him at the {\it Coll\`ege d'Harcourt}~\cite{wilson3}.}. According to Ballstadt, mathematics was the only branch
of 18th century science in which Diderot can be said to have been a practicioner~\cite{ballstadt2}.
So, if one considers Diderot's standing for the {\it Si\`ecle des Lumi\`eres} and his intellectual acumen, it is more than justifiable
that his mathematical exploits deserve a more detailed technical analysis.  

Thus, it is the purpose of this and the forthcoming  articles to expand some of the previous works by analysing Diderot's
works both from a more technical perspective while simultaneously highlighting the historical context of their writing.
In the present work his fifth memoir on the damped harmonic oscillator is discussed.
Besides presenting his calculations in modern notation, his results are compared to what we know today about the
damped pendulum. It is shown that if one considers the regime of high Reynolds numbers, Diderot's assumption of a damping force of
the type $F(v)\sim v^2$ is correct.
 %Moreover, there exists a notable similarity between these two memoirs of Diderot and
%the problems discussed by Huygens in his {\it Horologicum Oscillatorium}: the construction of an isochronous pendulum clockwhich requires
%the use of involutes, a concept introduced by Huygens in mathematics (for a definition, see below).

%{\it Memoirs on Different Subjects of Mathematics}. In the present
%work his fifth memoir on the damped harmonic oscillator is discussed.
%Besides presenting his calculations in modern notation, I compare his results to what we know today about the damped pendulum and show that if one considers
%the regime of high Reynolds numbers, Diderot's assumption of a damping force of the type $F(v)\sim v^2$ (where $v$ is the velocity of the bob) is not entirely
%devoid of meaning, as Coolidge argues~\cite{coolidge}. %Moreover, there exists a notable similarity between these two memoirs of Diderot and
%the problems discussed by Huygens in his {\it Horologicum Oscillatorium}: the construction of an isochronous pendulum clockwhich requires
%the use of involutes, a concept introduced by Huygens in mathematics (for a definition, see below).

This work is organized as follows. Section \ref{sec:DiderotMathematician} contains a brief description of the known mathematical works
of Diderot. This is followed by Section \ref{sec:DiderotMemoir}, where the historical context of Diderot's
Fifth Memoir, the main subject of this article, is discussed. The memoir itself is treated in Section  \ref{sec:DiderotPerspective}.
It starts by introducing Newton's discussion of the colliding pendulums and his assertions. Then, Diderot's own solution is presented
in all its mathematical detail. In order to make Diderot's calculations more transparent to the modern reader, his notation is explained
and some misprints in the original are pointed out. Any discussion of Diderot's (or, for that matter, Newton's) solution can only be appreciated if
one realizes the complexity of the problem they were dealing with. So, Section \ref{sec:MathematicalPendulum} gives a detailed treatment
of the pendulum for arbitrary swing amplitudes as well as the small-amplitude approximation, since Diderot considers both cases.
The effect of air resistance on the pendulum's movement is also discussed and an approximate solution using a
Lindstedt-Poincar\'e expansion is presented. Experimental support for a $v^2$-type drag is presented.
This section can be read independently and may be skipped by those interested only in the Diderot's solution. Some of the results
presented in this section are important for a better evaluation of Diderot's Memoir, but they are clearly pointed out along the text.
The paper closes with Section~\ref{Conclusions} where some conclusions are drawn.

\section{Diderot and Mathematics}
\label{sec:DiderotMathematician}
%\subsection{The memoirs of 1748, 1761 and Miscellanea}
In 1748, Pissot and Durant of Paris published an octavo volume with the unassuming title
{\it M\'emoires sur diff\'erens sujets de Math\'ematiques}~\cite{diderot1}. With its deluxe format and exquisite engravings, it
was {\it ``... one of the most coquettish [volumes] that was ever published on such arid subjects''}, as Maurice Tourneaux remarked~\footnote{Torneaux
and Jules Ass\'ezat were the first publishers of Diderot's complete works.}~\cite{wilson2}. The book contained five different treatises on different
subjects of pure and applied mathematics:

\begin{itemize}
 \item[I.] The first memoir is a study entitled {\it Principes g\'en\'eraux d'acoustique}. As the title indicates, it deals with acoustics and how one can
 relate the vibration of chords with particular musical notes, among other things. Diderot starts out with general properties
 of soundwaves and moves to the mathematics of pitch, and the use of logarithms in the production of harmonious sounds.
 For Coolidge this is the most important of all
 five memoirs~\cite{coolidge}. It is also the longest.
 \item[II.] The memoir {\it Examen de la d\'eveloppante du cercle} is a treatise on involutes. An involute
 is the curve obtained by the free end of a taut  string attached to a curved body as it unwinds from that body, as in the case of the spiral
 described by the tip of a rope as one unwrapps it from around a circle~\footnote{In the specialized
 mathematical literature involutes are also called `evolvents'.}.
 This work is particularly interesting for various reasons. In spite of reading like a piece of pure mathematics,
 Diderot never loses sight of applications. He starts the memoir by posing a practical question: whether it
 would be possible to draw curves without recurring to a ruler and a compass. In other words, he was looking for some kind of device with which one could
 draw `mechanical curves' ({\it courbes m\'ecaniques}) or curves that can be drawn with the help of some mechanical device\footnote{This term is no longer used and
 nowadays one speaks of algebraic curves (geometric curves in Diderot's language) and transcendental curves. Algebraic curves can be defined as the set of
 points given by the equation $f(x,y)=0$ where $f(x,y)$ is polynomial in $x$ and $y$. Transcendental curves intercept some straight line in an infinite number
 of points and cannot be represented by a polynomial equation of finite degree.}. More importantly, in this memoir he
 takes up on his hobbyhorse, the squaring of the circle, since involutes are intimately connected with the retification problem:
 given some region of space delimited by known curves, one expects to find its area by transforming these curves into straight ones with the help of involutes.
 Involutes were introduced by Christiaan Huygens (1629 -- 1695) in his treatise on pendulums and their applications in clockmaking,
 the {\it Horologicum Oscillatorium} of 1673 ~\cite{huygenswork}. They are also relevant in the design of mechanical gears, since dents which involute profiles have a
 better distribution of forces and are less prone to noise and wear, as first noted by Leonhard Euler
 (1707 -- 1783)~\cite{kk}. There is no indication whether Diderot knew of these works. He did however make use of the annotated edition of the {\it Principia}
 by Le Seur and Jacquier~\cite{jacsu}, where involutes are mentioned. See below and~\cite{dahmen2} for more details.
 \item[III.] In the third Memoir {\it Examen d'un principe de m\'ecanique sur tensions des cordes} one finds an experiment proposed by Diderot to decide a question
 posed by the italian mathematician Giovanni Borelli (1608 -- 1679): imagine a rope whose one end is attached to a fixed
 point, while a weight $A$ hangs at the other. Will the replacement of the fixed point by an equal weight change the resulting tension?
 The answer which today one may find in any elementary physics book was, at the time, an open question~\cite{kk}.
 \item[IV.] The fourth Memoir, {\it Projet d'un novel orgue} is related to the first memoir. Here Diderot introduces a project for an organ that could be played even by those who have no musical
 training and was based on the use of a sort of  punch card. This work had been published separately the year before in the {\it Mercure de France}~\cite{kk}.
\item[V.] The last memoir is entitled {\it Lettre sur la R\'esistance de l'air au mouvement des pendules}. This memoir is the subject of the present
article and as such will be discussed in more detail in Section \ref{sec:DiderotMemoir}.
 \end{itemize}
In 1761 Diderot wrote three more essays under the title {\it Noveaux M\'emoires sur  Diff\'erents Sujets de Math\'ematiques}, but which were kept private and
published only posthumously~\cite{diderot3}:
\begin{itemize} 
 \item[i.] An article on the cohesion of bodies. From today's
   viewpoint outdated, Diderot comes to the defense of Newton, arguing
   that an inverse-square law would suffice to explain cohesion
 \item[ii.] An article on the use of probability in calculating betting odds in the famous Saint Petersburg problem. In this article he corrected
 an error commited by D'Alembert while trying to solve that same problem. 
 \item[iii.] Another article related to probability, but this time on the question of inoculation. It was once again a response
 to D'Alembert's criticism, who thought very sceptically about the use of probabilities in matters of life and death.
 Diderot's involvement with `political arithmetics' or probability theory was much in line with his engagement on public issues and
 was more philosophical than mathematical in nature.
\end{itemize}

Further works, most in fragmentary form, involve questions as varied as were Diderot's interests.
There is a long article on cyclometry (squaring of the circle), which precedes a project of a deciphering machine; discussions on the
geometry of infinity and calculus of annuities (for insurance purposes); comments on celestial mechanics, the duration of human life and the outline of
a lottery; there is also a manual of basic arithmetics for children, as Diderot earned a living in his first years in Paris
by teaching mathematics to the children of well-off families. The most up-to-date edition of his extant mathematical works can be found in
the extensively annotaded {\it Oeuvres Compl\`etes} of 1975~\cite{diderot3}. 
\section{The Fifth Memoir on Newton and Colliding Pendulums}
\label{sec:DiderotMemoir}
 
Judged by its title, the fifth and last memoir of 1748 seems to be one of those paradigmatical exercises to be found in most physics textbooks: to
determine the retardation on the movement of a pendulum caused by air resistance . A more careful look at its content however, reveals that Diderot
does not seem to have been only interested in the problem {\it per se}, but also in giving a didactic explanation of a commentary made
by Newton in the first pages of the {\it Principia}, which Diderot goes as far as transcribing from in the original Latin
~\footnote{Diderot was known to be an accomplished latinist. He said he learned English by
translating works in this language into French with the help of a English-Latin dictionary~\cite{wilson4}.}.

The passage he quotes is concerned with is the experimental verification that Action $=$ Reaction, or the Third Law of Motion.
This problem, which at a first might look rather far removed from the problem of damped oscillators,  is actually the key to
Diderot's  assessment of Newton:
to verify the validity of the Third Law, Newton conducted a series of collision experiment with two pendulums. {\it `But'}, as Newton observes,
{\it `to bring this experiment to an accurate agreement with the theory, we are to have due
regard as well to the resistance of the air as to the elastic force of the concurring bodies'~
\footnote{ Verum, ut hoc experimentum cum theoriis ad amuffim congruat, habenda est ratio, cum restitentiae aeris,
tum etiam vis elasticae concurrentium corporum~\cite{principia}.}}. Newton however did not
bother to say how exactly the air resistance should be, except that the retardation was proportional to the arc of the trajectory.
By translating Newton's arguments into a more rigorous mathematical
language, Diderot showed that this assumption was equivalent to Newton having implicitly assumed a $v$-type law when in fact -- so thought Diderot --
he should have favoured a $v^2$-law. The way Diderot treated the problem and organized his article around Newton's ideas shows how competent
he was in his assessment of the great master. Diderot knew the didatic value of mathematics and  while filling in the gaps Newton left open,
he recast the problem in the language of differential calculus, translating Newton's geometric language
into a differential one. 

The motivation behind Diderot's study might have been a personal one: if one takes his dedicatory introduction at face value
\footnote{The introdution is dedicated to {\it M***}, whose identity is unknown.
The whole volume of Memoirs is dedicated most probably to Marie Anne Victoire Pigeon d'Osangis (1724 -- 1767), a french mathematician known
by the name of Madame de Pr\'emontval, as she was the wife of Pierre Le Guay de Pr\'emontval (1716 -- 1764), also a mathematician. {\it M***}
could have been just a fictive addressee as this Memoir in written in the form of a letter
~\cite{mayer}.},
he was asked to clarify a passage in the {\it Principia}:
 \begin{quote}
  {\it If the place [in the Principia] where Newton calculates the resistance caused by air on the movement of the pendulum embarasses you,
  do not let your self-steem be afflicted by it. As the greatest geometers will tell you, one encounters, in the depth and laconicity of the
  Principia, [enough] motives to completely console a man of penetrating mind who had some difficulty in understand them; and you will see
  shortly that there is another reason that seems even better to me -- that the hypothesis this author started with might not be exact.
  %Something surprises me however: that you were
  %advised to seek me in order to free you from your embarassment. It is true that I studied Newton with the purpose of elucidating him. I should
  %even tell you that this work was pushed on, if not sucessfully, at least with great gusto. But I did not think of it any longer since the Reverend Fathers
  %Le Sueur and Jacquier made their commentaries public, and I did not feel tempted to ever reconsider it. There was, in my work, a few things you would not
  %find in the work of this great geometers and a great many things in theirs you most surely would not find in mine. What do you ask of me?
  %Even though mathematical matters were once much familiar to me, to ask me now about Newton is to talk of a past dream. However, to persevere
  %in the habit of pleasing you I will leaf through my abandoned drafts, I will consult the sagacity of my friends and tell you what I can learn
  %from them, telling you also, with Horace: if you can make these better, please let me know. If not, follow them with me.
  \footnote{Si l'endroit \`ou Newton calcule la r\'esistance que l'air fait au mouvement d'un pendule vous embarasse, que votre amour-prope n'en
  soit point afflig\'e. Il y a, vous diront les plus grands g\'eom\'etres, dan s la profondeur et lat laconicit\'e des Principes math\'ematiques,
  de quoi consoler partout un homme p\'en\'etrant qui aurait quelque peine \`a entendre; et vous verrez bient\^ot que vous avez ici pour vous une
  autre raison que me para\^{\i}t encore meilleure; d'est que l'hypoth\`ese d'o\`u cet auteur est parti n'est peut-\^etre pas exacte~
  \cite{diderot1}.}}
 \end{quote}
 \vskip 0.5cm
 %
%He is also very humble in saying that
%the last word on the controversy should come from experiments~\cite{diderot1}:
% \begin{quote}
%  {\it Fifth experiment: To ascertain whether the retardation that air causes on the movement of pendulums are as
%  the arcs or the square of the arcs and to repeat Newton's experiments on the collision of bodies~\footnote{S'assurer si les retardations que
%  l´air fait au mouvement des pendules sont comme les arcs ou comme les carr\'es des arcs, et recommencer les
%  exp\'eriences de Newton sur le choc des corps.}.}
% \end{quote}
% \vskip 0.5cm
 \noindent
 The error Diderot is talking about is Newton's choice of a force linear in $v$, which he believes should be quadratic.
 This is the reason why previous works on the subject have given emphasis to the $v$ vs. $v^2$ controversy~\cite{kk,coolidge},
 when the truth is that Newton actually considered both types of force in the {\it Principia}. The first 31 propositions of Book II
 are dedicated to the problem of damped pendulums, as discussed extensively in a series of articles by Gauld~\cite{gauld1,gauld2}.
 Actually Newton went so far as to use the more general expression $F_R(v)= a\;v + b\;v^{\frac{3}{2}} + c\;v^2$ in order to fit the results
 of experiments he conducted himself. At the end of Book II, Section I of the  he affirms by way of conclusion~\cite{principia3}:
 \begin{quote}
  {\it However, that the resistance of bodies is in the ratio of the velocity, is more a mathematical hypothesis than a physical one. In mediums
  void of all tenacity, the resistance made to bodies are as the square of the velocities. For by the action of a swifter body, a greater motion
  in proportion to a greater velocity is communicated to the same quantity of the medium in a less time; an in an equal time, by reason of a
  greater quantity of the disturbed medium, a motion is communicated as the square of the ratio greater; and the resistance (by Laws II and III)
  is as the motion communicated.}
 \end{quote}
Not surprisingly this explains why, in the specialized literature on friction, a $v^2$-dependent $F_R$ is known as {\it Newton Friction},
whereas a $v$-dependent $F_R$ is called {\it Stokes Friction}.

Newton's approach to the damping problem was criticized by Leonhard Euler and Daniel Bernoulli (1700--1782) for its lack of rigour, something which
certainly did not baffle a practical mind like Diderot's as much as it baffled those of the great hydrodynamicists. Diderot's respect for Newton
was too great: {\it `I have for Newton all deference one accords to the unique men of his kind'}\footnote{J'ai pour Newton toute la d\'ef\'erence
qu'on doit aux hommes unique dans leur genre~\cite{diderot1}}.
So Diderot might have been motivated
by something other than someone's request: he seems to have had the intention of publishing his own commentaries on the
{\it Principia}, but he was superseded by the famous annotated edition of the Franciscan Fathers Fran{\c c}ois Jacqueur (1711--1788) and
Thomas Le Sueur (1703--1770) which came out between 1739 and 1742. So we read from his introduction:
\begin{quote}
{\it
Something surprises me however: that you were
advised to seek me in order to free you from your embarassment. It is true that I studied Newton with the purpose of elucidating him. I should
even tell you that this work was pushed on, if not sucessfully, at least with great vivacity. But I did not think of it any longer since the Reverend Fathers
Le Sueur and Jacquier made their commentaries public, and I did not feel tempted to ever reconsider it. There was, in my work, a few things you would not
find in the work of these great geometers and a great many things in theirs you most surely would not find in mine. What do you ask of me?
Even though mathematical matters were once much familiar to me, to ask me now about Newton is to talk of a dream of a year past. However, to persevere
in the habit of pleasing you I will leaf through my abandoned drafts, I will consult the lighra of my friends and tell you what I can learn
from them, telling you also, with Horace: if you can make these better, please let me know. If not, follow them with me
\footnote{Mais une chose me surprend; c'est que vous vous soyez avis\'e de vous adresser \`a moi, pour vous tirer d'embarass. Is est vrai que
j'ai \'etudi\'e Newton, dans le bassein de l'\'eclaircir; je vous avouerai m\^eme que ce travail avait \'et\'e pouss\'e, sinon avec
beaucoup de succ\`es, du moins avec assez de vivacit\'e; mais je n'y pensais plus d\`es le temps que les RR P\`eres Le Sueur et Jacquier donn\`erent
leur Commentaire; et je n'ai point \'et\'e tent\'e de le rependre. Il y aurait eu, dans mon ouvrage, fort peu des choses qui ne soient dans
celui des savants g\'eom\`etres; et il y en a tant dans le leur, qu'assur\'ement on n'e\^ut pas rencontr\'ees dans le mien! Qu'exigez-vous de
moi? Quand les sujets math\'ematiques m'auraient \'et\'e jadis tr\`es-familiers, m'interroger aujourd'hui sur Newton, c'est me parler
d'un r\^eve de l'an pass\'e. Cependant, pour pers\'ev\'erer dans l'habitude des vous satisfaire, je vais, \`a tout hazard, feuilleter mes
paperasses abandonn\'ees, consultes les lumi\`eres de mes amis, vous communiquer ce que j'en pourrai tirer, et vous dire, avec Horace: Si quid
novisti rectius istis, candidus imperti. Si non, his utere mecum
\cite{diderot1}.}.}
\end{quote}

The fifth memoir might have also been part of a more general work: given that Diderot also wrote an article
on involutes (Second Memoir) and these are intrinsically connected with the problem of constructing an isochronous pendulum, these two memoirs
might bear some relation with Christiaan Huygens epochal {\it Horologium Oscillatorium}~\cite{huygenswork}. It would well fit the interests
of Diderot in the `applied arts' and the fact that earlier in his career he prepared the general formulas and mathematical tables for a
treatise of Antoine Deparcieux (1703 -- 1768) on sundials~\footnote{It is thus not surprising that the entry {\it Cadran Solaire} (Sundial)
in the {\it Encyclop\'edie} was signed by him and D'Alembert.}. Timekeeping devices could have exerted a certain fascination
on him~\cite{wilson5}.
Tempting as this supposition
might be, there is to the author's knowledge no mention of Huygens in Diderot's works. This does not mean that he did not know it,
as Diderot was very well acquainted with the mathematical literature of his times. The memoirs of 1748 themselves might also have been
an attempt of the author to appear more serious to the eyes of his contemporaries: by that year he had acquired a rather scandalous
reputation with the publication of a rather brazen novel entitled {\it Les Bijoux Indiscrets}. It was published anonymously, but
the author's name was no secret. The strongest evidence of Diderot's attempt to look serious can be
read off from the opening phrase of his Memoirs, drawn from Horace's Satires: `{\it [Sed tamen] amoto quaeramus seria ludo}', which
roughly translates as `plays aside, let us turn to serious matters.' Irrespective of his ultimate motivation, his Fifth Memoir is the
embodiment of his competence in things mathematical.

%======
\section{The Mathematical Pendulum from Diderot's Perspective}
\label{sec:DiderotPerspective}

To understand Diderot's approach, one has to first consider Newton's experiment described in the initial pages of the {\it Principia}. This is
the more so if one realizes that Diderot wrote his article as some sort of solution's manual to the arguments Newton expounded. The path Diderot chose
to arrive at answers to the questions he poses at the beginning of his memoir seems at first rather awkward. But once one realizes that he is following
Newton's logics closely, translating his arguments into mathematical form, one understands why he chose to solve the problem the way he did.

\subsection{Newton's Solution}
\label{sec:NewtonSolution}
Newton's discussion is based on the experimental setup depicted in Fig.~\ref{newtonpendulum} below. It is reproduced in the section {\it Axioms, or
Laws of Motion} of the {\it Principia}. It was inspired on earlier experiments done by Edme Mariotte (1620--1684) and Christopher Wren (1632--1723)
on the collision of pendulums. 
\begin{figure}[h]
\centering
\includegraphics[scale=0.5]{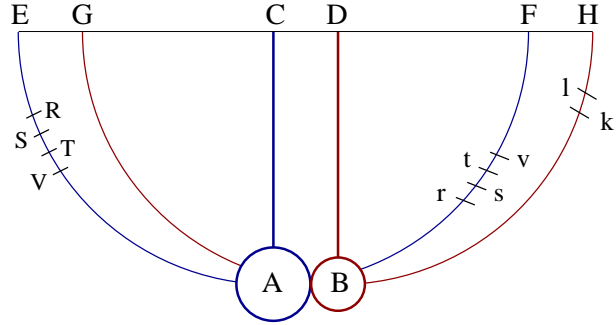}
 \caption{The pendulum Newton considered in his experiment to prove that Action equals Reaction: As bob $A$ is let loose
 from a given point $R$, it hits $B$. As a consequence of their collision, both move upwards, $A$ reaching $s$ and $B$ reaching $k$. The different points
 of the trajectories marked $r$, $t$, $v$ etc. and what they represented are discussed in the text. Diderot reproduced this picture in his memoir on the
 damped pendulum.}
 \label{newtonpendulum}
\end{figure}

\label{DiderotPendulum}
%===
Newton wants to study the transfer of momentum between colliding pendulums. The collision happens at point $A$. 
Given that the bob will have a lower velocity at $A$ as compared to what it would in vaccuum,
one has to correct for the lost momentum. To find this, Newton devises a simple trick:
since this difference is proportional to the
path traversed (see Fig. \ref{newtonpendulum}), {\it `...For it is a proposition well known to geometers, that the velocity of a pendulous
body in the lowest point is as the chord of the arc which it has described in its descent'
~\footnote{Nam velocitatem penduli in puncta infimo esse ut
chordam arcus, quem cadendo descripsit, propositio est geometris notissima~\cite{principia}.}}, after a full swing the pendulum will return to
point $V$ so that $RV$ represents the full retardation. As one complete oscillation
is made up of four quarter oscillations, and the four retardations are increasingly smaller, one has to determine how much the first retardation
contributes to the full $RV$. The easiest solution is to say that all contribute the same amount $(1/4) RV$. However, if one wants to minimize his
error, one can devise a trick: there must exist
a point $S$ below $R$ so that the momentum the bob loses upon reaching $A$ will be exactly $RS = RA-SA=1/4 RV$. If one is able to find this $S$,
then one can be sure that a bob starting from that point will lose momentum which corresponds exactly to a retardation $(1/4) RV$. 
Newton knows that $S$ should actually be placed a little further down, but not too much.
So he chooses a point $T$ in order to constrain how far down he can place $S$ by making $ST= 1/4 RV$. He then
places $ST$ halfway between the observed values of $R$ and $V$ (see Fig. ~\ref{newtonpendulum}). Newton's original passage reads~\cite{principia}:
\begin{quote}
 {\it Let the spherical bodies $A$, $B$ be suspended by equal and parallel strings $AC$, $BD$ from centers $C$ and $D$.
 About these centers and lenghts they describe  the semicircles $EAF$, $GBH$ with at $CA$ and $DB$.
 Bring the body $A$ to any point $R$ of the arc $EAF$, and (withdrawing the body $B$ let it go from thence, and after one oscillation
 suppose it to return to the point $V$: then $RV$ will be the retardation arising from the resistance of the air. Of this $RV$ let $ST$ be the
 fourth part, situated in the middle, namely so that
 \begin{equation}
 RS=TV
 \end{equation}
 and
\begin{equation}
 RS:TV = 3:2
\end{equation}
then will $ST$ represent very nearly the retardation during the descent from $S$ to $A$
\footnote{Pendeant corpora sphaerica A, B filis parallelis et aequalibus AC, BD, a centris C, D. His centris et intervallis describantur
semicirculi EAF, GBH radiis
CA, DB bisecti. Trahatur corpus A
ad arcus I E A F punctum quodvis R, et (subducto corpore B) demittatur inde, redeatque post unam oscillationem ad punctum
V. Est R V retardatio ex resistentia aeris. Huius R V fiat S T pars quarta sita in medio, ita scilicet ut R S et
TV aequentur, sitque R S ad ST ut 3 ad 2. Et ista ST exhibebit retardationem in descensu ab S ad A quam proxime \cite{principia}.}.}
\end{quote}

There does not seem to be any particular good reason for choosing this point other than the fact that he knows that the quarter retardations
are not equal and the first one is largest. So instead of choosing
\begin{equation}
 S= R - \frac{1}{4} RV
\end{equation}
he chooses
\begin{equation}
\label{eq:newtonvalueofs}
 S = R - \biggl( \frac{1}{4} RV + \frac{1}{8} RV  \biggr)
\end{equation}
He then proceeds, considering what happens at the other side, where the bobs ascend:
\begin{quote}
{\it
 Restore the body $B$ to its place: and supposing the body $A$ to be let fall from the point $S$, the velocity thereof in the place of
 reflection [i.e. collision] $A$, without sensible erros, will be the same as if it had descended {\it in vacuo} from the point $T$ ... After reflection,
 suppose the body $A$ comes to the place $s$ and the body $B$ to the place $k$. Withdraw the body $B$, and find the place $v$, from which,
 if the body $A$, being let go, should after one oscillation return to the place $r$, $st$ may be a fourth part part of $rv$, so placed in
 the middle thereof as to leave $rs$ equal to $tv$, and let the chord of the arc $tA$ represent the velocity which the body $A$ had in the
 place $A$ immediately after reflection. For $t$ will be the true and correct place to which the body $A$ should have ascended, if the resistance
 of air had been taken off. In the same way wer are to correct the place $k$ to which the body $B$ ascends, by finding the place $l$ to which
 it would have ascended {\it in vacuo}. And thus everything may be subjected to experiment, in the same manner as if we were really placed
 \it in vacuo~\footnote{Restituatur corpus
B in locum suum. Cadat corpus A de puncto S, et velocitas eius in loco reflexionis A sine
errore sensibili tanta erit, ac si in vacuo cecidisset de loco T ... Post
reflexionem perveniat corpus A ad locum s, et corpus B ad locum k. Tollatur corpus B et inveniatur locus v; a quo si corpus A
demittatur et post unam oscillationem redeat ad locum r, sit st
pars quarta ipsius rv sita in medio, ita videlicet ut rs
et $tv$ baequentur; et per chordam arcus $tA$
exponatur velocita s, quam corpus A proxime post reflexionem habuit in loco A. Nam t
erit locus ille verus et correctus, ad quem corpus A,
sublata aeris resistentia, ascendere debuisset. Simili methodo corrigendus erit locus k,
ad quem corpus B ascendit, et inventendus locus l, ad quem corpus illud ascendere debuisset in vacuo~\cite{principia}.}.}
\end{quote}
So, by using similar procedures for the ascent of the bobs, he is able to correct for their momentum. His conclusion~\cite{principia}:
\begin{quote}
 {\it Thus trying the thing with pendulums of 10 feet, in unequal as well as equal bodies, and making bodies concur after a descent through
 large spaces, as of 8, 12 or 16 feet, I found always, without an error of 3 inches, that when bodies concurred together directly, equal changes
 towards the contrary parts were produced in their motions, and, of consequence, that the action and reaction were always equal
 \footnote{
 Hoc modo in pendulis pedum decem rem tentando, idque in corporibus tam inaequali
bus quam aequalibus, et faciendo ut corpora
de intervallis amplissimis, puta pedum octo
vel duodecim vel sexdecim, concurrerent;
repperi semper sine errore trium digitorum
in mensuris, ubi corpora sibi mutuo directe
occurrebant, aequales esse mutationes motuum
corporibus in partes contrarias illatae,
atque ideo actionem et reactionem semper esse aequales
 ~\cite{principia}.}.}
\end{quote}

This are the last words of Newton that Diderot transcribes in his Memoir. The he goes on to determine, in a rigorous way, the location of $S$
under the assumption of an air resistance force linear in $v$.

\subsection{Diderot's Solution: Lettre sur la R\'esistance de l'Air}
Since Diderot's article aims at explaining how Newton got his $S$ while at the same time changing the hypothesis as to what regards the kind of
drag one uses, his Memoir is didactically organized in three mains parts:
two Propositions and the {\it \'Eclaircissements} (Clarifications). The Propositions deal with ways of calculating the retardation in the case of
a quadratic resistance force. The presentation, which might at first seem quite awkward to the reader, follows closely Newton's logic except
that Diderot uses a different hypothesis and makes extensive use of differential tools. The {\it \'Eclaircissements} are the place
where Diderot actually solves for Newton's $S$. Rather surprising is the fact that the {\it \'Eclaircissements} can be read quite independently,
since Diderot makes no direct use of his results from Proposition I and II. So one may rightfully ask the reason why he goes at pains to do all
the calculations which he does not use at the end. In the authors opinion this was part of his strategy: other than simply offering the solution to
a problem posed, Diderot shows in the {\it \'Eclaircissements} that Newton's answer is an approximation to the full solution which would follow
as a direct consequence of the {\it methodological} approach that he, Diderot, developed. If his reasoning, mathematically formulated, allowed him
to go beyond Newton, then his solutions of Propositions I and II must be correct. The emphasis is on the {\it method}, not on the {\it solution}

\begin{figure}[h]
\centering
\includegraphics[scale=0.5]{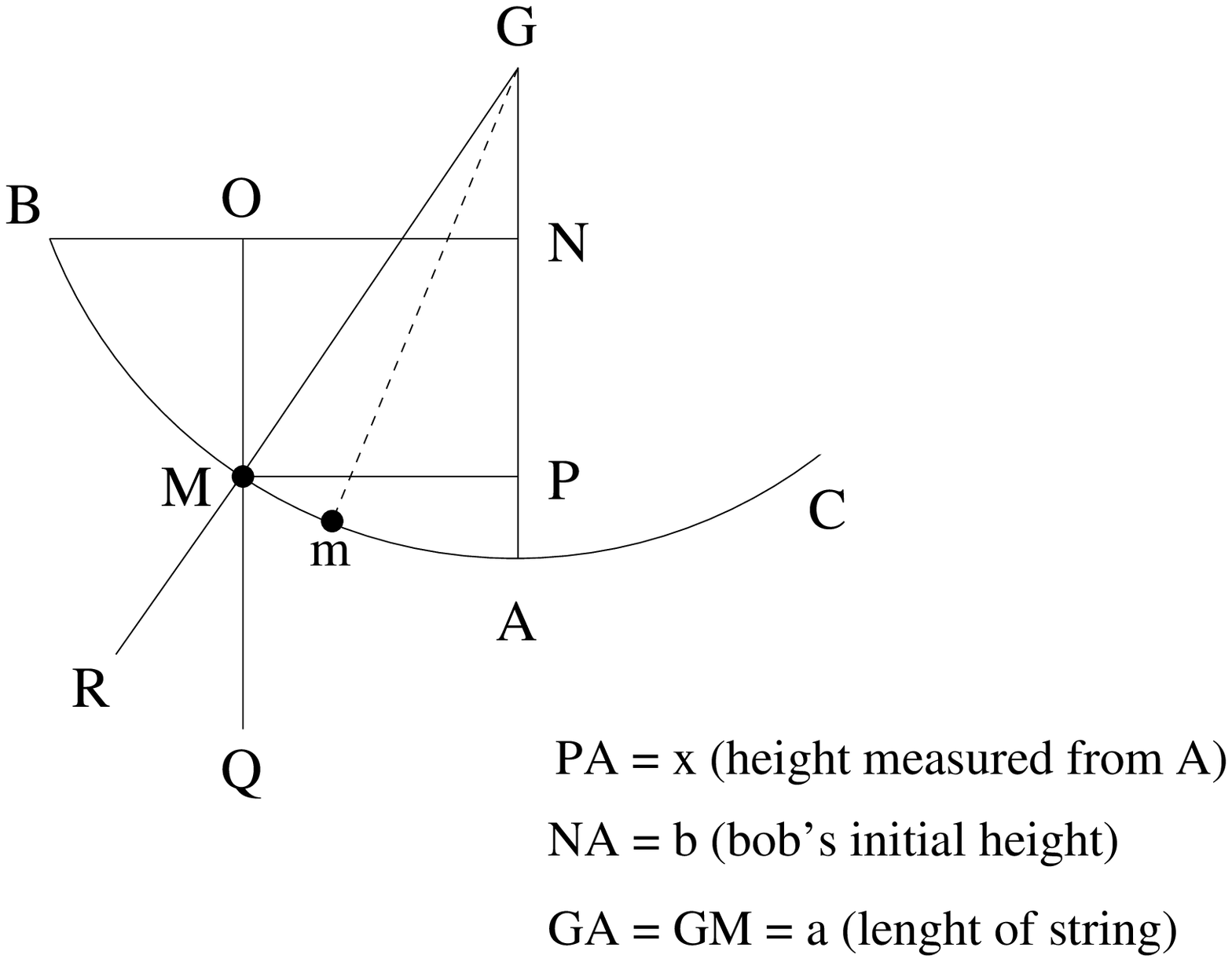}
 \caption{The pendulum in Diderot's work: $M$ represents an arbitrary point of the trajectory of a bob dropping from the initial position $B$. The position $m$ is infinitesimally close
  to $M$. Diderot expresses the position of the bob in terms of the height $x$ of point $M$, measured relative to the lowest point of the trajectory, that is,
  the segment $AP$ (per definition height at $A$ is $x=0$). The initial height is
  $b= NA$. The length of the pendulum is $a = GA = GM$.}
  \label{diderotpendulumfig1}
\end{figure}

Thus, he starts each proposition in the form of a homework, a {\it Probl\'eme} that he poses: to find the velocity $v$ of a bob for an arbitrary point $M$ along the
trajectory given that besides the weight, the bob is also acted upon by a retarding force proporcional to $v^2$.  Proposition I deals with
the bob's way from $B$
(to the left of the vertical $OA$) as it moves down to $A$, the lowest point of the trajectory (see Fig.~\ref{diderotpendulumfig1}).
Proposition II deals with the movement of the bob initially at $A$ as it moves up towards the right after being given an
initial velocity $h$. The separation of the question into two separate ones is due to the fact that Newton discusses each quarter
cycle independently. This is natural in the context of Newton's commentaries: since Newton was interested in the collision of two bobs,
the descending bob will execute a quarter of a cycle before colliding.  After Diderot gives a  {\it Solution} to a {\it Probl\'eme}, he writes down a
few extra corollaries, which are either straightforward consequences of his main solution or approximations that one gets when considering small
angles of oscillation.
\vskip 0.5cm
\noindent
{\bf Proposition I:} {\it Let a pendulum $A$ which describes an arc $BA$ in air be attached to the string $GM$ fixed at $G$. One asks for the velocity
of this pendulum for any point $M$, assuming that it starts falling from point $B$~\footnote{Soit un pendule $M$ qui d\'ecrit dans l'air l'arc $BA$,
\'etant attach\'e \`a la verge $GM$ fixe en $G$. On demande la vitesse de ce pendule en un point quelconque $M$, en supposant qu'il commence
\`a tomber du point $B$~\cite{diderot1}.}}

Before we discuss Diderot's solution, his choice of variable requires some explaining: instead of using $\theta$, the displacement angle,
as one would normally do nowadays, he prefers to think in terms of the
height $x$ of the bob relative to the lowest point $A$  of the trajectory. There is a reason for this: in the absence of damping, by conservation of energy
we know that the change in kinetic energy of the bob is equal to change in potential energy. This allows one to directly find the velocity
at a given height $x_1$
by giving the difference in height $x_{0}-x_1$ through which the bob of mass $m$ fell, that is
\begin{equation}
 \frac{1}{2}m{v_1}^2-\frac{1}{2}m{v_0}^2 = mg(x_0-x_1)\longrightarrow {v_1}^2={v_0}^2 + 2 g (x_0-x_1).
\end{equation}
This is Torricelli's equation for a body with accelaration $g$. Even though this result does not hold in the presence of damping,
one may still use it as a first approximation to the real velocity, as Diderot eventually did.

The height from which the bob starts is the orthogonal projection of point B on line OA, and this Diderot calls $b= x_{0}= x(t=0)$ (see Fig.
~\ref{diderotpendulumfig1}).
The length of his string is $a$ (usually called $l$ in modern texts). He assumes that the force due to air resistance is given by
 \begin{equation}
\label{eq:dragforcediderot}
F(v)= \gamma\;v^2
\end{equation}
which he writes as
\begin{equation}
F(v)=\frac{f}{g^2}\;v^2
\end{equation}
The modern reader might find this a bit confusing but this equation comes from the fact that Diderot assumes that for a given
known velocity $g$ (not to be confused with the acceleration of gravity)
the force has a known value of $f$. So, from (\ref{eq:dragforcediderot}) one has
\begin{equation}
F(g)= f= \gamma\;g^2\longrightarrow \gamma= \frac{f}{g^2}
\end{equation}
The factor $f/g^2$ is carried along through the whole text. The determination of this term is no simple experimental task and Diderot lacked access to scientific
apparatus. So, even if conceptually correct, the use of the factor $f/g^2$ might have served him the purpose of convincing his readers that the problem was real,
not just a toy model. For the sake of a more compact notation we 
will keep the parameter $\gamma$ where Diderot uses $f/g^2$ and think of its determination
as it is normally done in a laboratory experiment: by fitting the amplitude as it decays with time or by measuring the drag in a wind tunnel.

Diderot's approach consists in finding a relation between $dv$, the increment in velocity, and the difference in height $dx$ associated to the fall.
The equation of motion for a bob of weight $p$ acted upon by a force of the type Eq. (~\ref{eq:dragforcediderot}) is
\begin{equation}
m\frac{dv}{dt} = p\sin\theta - \gamma\;v^2
\end{equation}
To write it in terms of $dx$ Diderot needs to find a way to relate this to $dt$. He begin by noticing that
\begin{equation}
\label{eq:basic}
dt= \frac{ds}{v}
\end{equation}
where $ds$ (Diderot's arc $Mm$) is the length the bob traverses along the arc during the time interval $dt$. So, replacing $dt$ by this expression he gets
\begin{equation}
m\;dv = (p\sin\theta - \gamma\;v^2)\times \frac{ds}{v}
\end{equation}
or
\begin{equation}
\label{eq:retardedbobdown}
v\;dv = (p\sin\theta - \gamma\;v^2)\times \frac{ds}{m}
\end{equation}

\begin{figure}
\centering
\includegraphics[scale=0.5]{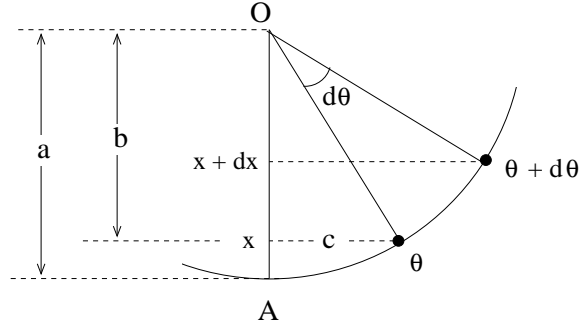}
\caption{The relation between $\theta$ and $x$, which Diderot uses as variable in his memoir. The angle $\theta$ is at the vertex
of a right triangle of sides $a$, $b$, and $c$. From these lenghts and a few trigonometric identities one can find the relation between
$\theta$ and $x$.}
\label{diderotpendulumfig2}
\end{figure}

In Diderot's original work the mass $m$ of the pendulum does not
appear. This could be a lapse, not a conceptual mistake, or the fact that Diderot took $m=1$ without mentioning it.
For the sake of completeness we will keep the mass $m$ in the equations that follow. 
To go over now to Diderot's $x$ one has to first remember that an infinitesimal arc $ds$ is related to the infinitesimal angular displacement $d\theta$ via
$ds = a\;d\theta$. The relation between $\theta$ and $x$ can be easily inferred from Fig.~\ref{diderotpendulumfig2} 
and some basic trigonometry:
\begin{equation}
\label{eq:thetaxrelation}
\sin\theta=\pm\;\frac{\sqrt{2ax - x^2}}{a}
\end{equation}
\noindent
The $\pm$ sign comes from the fact that the expression on the right-hand side is always positive but $\sin\theta$ can be either positive
or negative depending on the which side the bob is. In other words, for a bob moving from left to right one always has $d\theta > 0$ while
on the journey down $dx < 0$ while on the way up $dx > 0$. This is what Diderot means when he says
\begin{quote}
{\it In this equation I substitute the little arc
$Mm$ [ds] by its value $-\frac{a}{\sqrt{a^2 - b^2}}\;dx$, with a minus sign, because as the pendulum goes down the velocity increases while x becomes smaller.
~\footnote{Dans cette \'equation, je mets, au lieu du petit arc $Mm$ sa valeur  $-\frac{a}{\sqrt{a^2 - b^2}}\;dx$, avec le signe $-$,
parce que $v$ croissant \`a mesure que le pendule descend, $x$ diminue au contraire~\cite{diderot1}.}}
\end{quote}
Differentiating both sides of Eq. (\ref{eq:thetaxrelation}) one gets
\begin{equation} 
\cos\theta\;d\theta= \pm\;\frac{a-x}{a\sqrt{a^2 - b^2}}\;dx
\end{equation}
or remembering that $\cos\theta= \frac{a-x}{a}$ this can be recast as
\begin{equation}
d\theta= \pm\;\frac{dx}{\sqrt{a^2 - b^2}}
\end{equation}
from which one gets
\begin{equation}
v\;dv = p\;(-dx)  - \gamma\;v^2\times \frac{a(-dx)}{\sqrt{2ax - x^2}}
\end{equation}
which is how Diderot writes (\ref{eq:retardedbobdown}). Integrating both sides of this equation one ends up with
\begin{equation}
 \frac{v^2}{2} = -\frac{p}{m}\;\int_{b}^{x} dx' + \int_{b}^{x}\frac{\gamma}{m}\frac{v^2\;a\;dx'}{\sqrt{2ax - x^2}}
\end{equation}
or
\begin{equation}
\label{eq:diderotfirstequation}
 \frac{v^2}{2} = \frac{p}{m}(b-x) + \int_{b}^{x}\frac{\gamma}{m}\frac{v^2\;a\;dx'}{\sqrt{2ax - x^2}}
\end{equation}
Another point to note is that Diderot never writes explicitly the upper and lower limits of integration but does it by explicit comments in the text.
This is typical of him, who sometimes explains (when he does!) his steps in writing and not in formulas~\cite{kk}.

Diderot thus ends up with an integral equation for $v$, which he cannot solve. He notes however that, in the absence of air, the speed of a pendulum
falling from rest from $B$ to $M$, that is, from a height $b$ to a height $x$ is simply
\begin{equation}
\frac{mv^2}{2} = p(b-x)
\end{equation}
which follows from conservation of energy.
In order to handle Eq. (\ref{eq:diderotfirstequation}) Diderot uses the following argument: given that the drag is much smaller than the weight
of the bob, one may assume that `{\it $v^2$ diff\'erera tr\`es--peu de $2\;p\;b -2\;p\;x$'} [notice there is a mass $m$ missing in Diderot's calculation]. One may, therefore, substitute $v^2$ inside the integral
by its approximate value $(2\;p\;b -2\;p\;x)/m$ to finally write
\begin{equation}
\label{eq:diderotintegral}
 v^2 = 2\;\frac{p}{m}(b-x) + 2\int_{b}^{x}\frac{\gamma}{m^2}\frac{(2pb-2px')\;a\;dx'}{\sqrt{2ax' - x'^2}}
\end{equation}
What Diderot does, in modern parlance, is a first-order approximation, that is, to substitute for $v^2$ in the integral the value it would
have in vacuum and thus obtain a correction. A zeroth-order approximation would be to assume $v^2$ in air to be the same as in
vacuum.

In the solution of (\ref{eq:diderotintegral}) one can nicely see how Diderot was, foremost, a geometrician, but one who was already moving into a more analytical
approach, since he mixes geometrical ideas with analytical ones to solve the integral.

The first integral to be solved is
\begin{equation}
2\int_{b}^{x}\frac{\gamma}{m^2}\frac{2pb\;a\;dx'}{\sqrt{2ax' - x'^2}} = \frac{4\;p\;b\;\gamma}{m^2}\int_{b}^{x}\frac{a\;dx'}{\sqrt{2ax' - x'^2}}
\end{equation}
But Diderot knows, without bothering to say, that the integrand is just the infinitesimal arc $Mm$, so the integral is nothing but the arc measure from point $B$
to point $M$, that is, his $BM$. So he writes his answer as
\begin{equation}
\int\frac{b\;a\;dx'}{\sqrt{2ax' - x'^2}} = -b \times\; BM
\end{equation}
where the $-$ sign comes from his convention for the sign of $dx$. The remaining part of (\ref{eq:diderotintegral}) is a bit harder. Diderot writes down
\begin{equation}
\label{eq:diderotintegral2} 
\int\frac{-a\;x'\;dx'}{\sqrt{2ax' - x'^2}} = \int\frac{(a^2 -a\;x')\;dx'}{\sqrt{2ax' - x'^2}} - \int\frac{a^2\;dx'}{\sqrt{2ax' - x'^2}}
\end{equation}
This rewriting of the equation, by adding and subtracting the same term, is easily explained. Diderot knows that
\begin{equation}
a\frac{d}{dx}(\sqrt{2ax-x^2})=\frac{a^2-ax}{\sqrt{2ax-x^2}}
\end{equation}
and he can thus write (\ref{eq:diderotintegral2}) as
\begin{equation} 
\int\frac{-a\;x'\;dx'}{\sqrt{2ax' - x'^2}} = a\;\int\frac{d}{dx'}\sqrt{2ax' - x'^2}\;dx' - a\;\int\frac{a\;dx'}{\sqrt{2ax' - x'^2}}
\end{equation}
The solution of the first integral on the right hand side is trivial, because $\int_{a}^{b} (df/dx)dx = f(a)-f(b)$. Moreover, and here his geometrical
intuition comes to his help, the integrand $\sqrt{2ax - x^2}=\sqrt{a^2 -(a - x)^2}$ is just the distance from point $M$ to the vertical axis $OA$, that is, the straight
line $MP$. With can then finally bring all these results into one equation and write~\footnote{Diderot of course uses $f/g^2$ instead of our $\gamma$.
In the original article there is also a misprint: there is a $g^2$ factor missing in the denominator of the $(B0-BM)$ term.}
\begin{equation}
m v^2 = 2\;p\;(b-x) -\frac{4\;\gamma}{m}\;p\;b\times\;BM -\frac{4\;\gamma}{m}\;p\;a\times\; (BO-BM)
\end{equation}
which is the solution of his original question expressed in terms of the arc $BM$ and the distance $BO$.
In Diderot's variable $x$ this would read:
\begin{equation}
mv^2 = 2\;p\;(b-x) +\frac{4\;\gamma}{m}\;p\;a\;(a-b)\;\biggr[ \cos^{-1}\;(1-b/a) -\cos^{-1}\;(1-x/a) \biggr]
-\frac{4\;\gamma}{m}\;p\;a\times\;(\sqrt{2ab-b^2} -\sqrt{2ax-x^2})
\end{equation}
There is however a certain charm (and economy) in Diderot's original notation, because his equation allows one to
come up with a nice geometrical interpretation of velocity correction directly in terms of the distances traversed by the pendulum: $BM$ represents the distance
the bob travels along the circular arc and $BO$ measures how far it moves to the right, from $B$ to $M$. This allows one also to study some interesting limits, which
Diderot does in corollaries I, II and III. Corollary I is just the expression above calculated for $x=0$, that is, what the velocity looks like when the bob reaches
A. Corollary II is a rather trivial observation (but important for Diderot's subsequent discussions) and follows from a rewriting of the equation above: that
the velocity obtained is the same velocity of pendulum that falls without air resistance from a starting point below point $B$. He puts the problem always in
terms of comparisons between fall in vaccuum as opposed to fall in viscous medium in order to prepare the reader for his discussion of Newton.

Corollary III follows when considering what would happen if one had a small initial amplitude. In this case $BM \approx BO$. With $x=0$
Diderot's expression becomes
\begin{equation}
 mv^2 = 2\;pg\;b\;\biggl( 1 - \frac{2\;\gamma}{m} BM \biggr)
\end{equation}
which is Diderot's correction to Torricelli. With these 3 corollaries Diderot moves over the the second part of his problem: how to determine the
velocity of the bob on the way up, for an arbitrary point $M$, given an initial velocity at $A$ equal to $v^{(0)}_A$ (Diderot calls this initial velocity $h$). 
The part devoted to the second proposition is longer, not because the problem is more difficult -- what he has to do now is basically to revert the sign of $dx$
in the equation he already had and add an initial velocity $h$ -- but because in this section he derives the result that retardation goes as (arc)$^2$ and
not linear in the arc, as Newton assumed.
\vskip 0.5cm
\noindent
{\bf Proposition II.} {\it `Suppose that a pendulum $A$, placed initially at the vertical $GA$, is given an impulse or velocity $h$ along the horizontal $AR$.
One wants to know its velocity for an arbitrary point $M$.'~\footnote{Supposons qu'un pendule, plac\'e dans la situation vertical GA, re{\c c}oive une
impulsion ou vitesse h suivant l'horizontale AR. On demande sa vitesse en un point quelconque M~\cite{diderot1}.}}
\begin{figure}[h]
\centering
\includegraphics[scale=0.5]{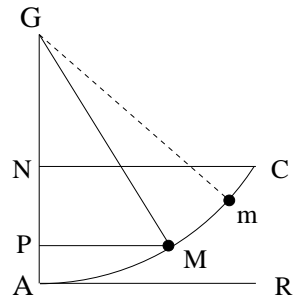}
\caption{The figure used by Diderot in his study of the movement of the bob upwards. As always, $Mm$ represents an infinitesimal displacement for Diderot.}
\label{diderotpendulumfig3} 
\end{figure}
\\
\noindent
Diderot repeats the same steps as before but making sure that in the new equation the signs of $dv$ and $dx$ are opposite, as the velocity decreases as the
bob moves upward (see Fig. ~\ref{diderotpendulumfig3}).
Once again he uses Torricelli's equation
\begin{equation}
 v^2 = (v^{(0)}_A)^2 +2\;a\;(x_{final} - x_0)= h^2 - 2\frac{p}{m}x
\end{equation}
for an initial velocity $v^{(0)}_A = h$ and acceleration $a=p/m$ to solve the integral in approximate form.
As his calculations are basically the same, we will not repeat them. He arrives at the following answer
\begin{equation}
m v^2 = m h^2 -2\;p\;x-\frac{2\;\gamma}{m}\;h^2\times\;AM - -\frac{4\;\gamma}{m}\;p\;a\times\; (AM-MP)
\end{equation}
He now takes a different path. Instead of leaving the equation as it is, he substitutes $h^2$ by the maximum height $AN$ the pendulum would reach in vacuum. 
This can be easily done since the highest point is where $v=0$. So, if one takes the expression above for $v=0$ and $\gamma=0$ he finds
\begin{equation}
 m\;h^2 = 2\;p\times AN
\end{equation}
Substituing this value back into his answer and noticing that $(AN -x) = NP$, he writes
\begin{equation}
m v^2 = 2\;p\times\;NP -\frac{4\;\gamma\; p}{m}\;\times\;AM\times AN +\frac{4\;\gamma\;p}{m}\times\;a\times\; (AM-MP)
\end{equation}

This expression is the starting point for his corollary I of proposition II, namely,  to find the highest point reached by the pendulum in the presence of
drag. This can be easily obtained by setting $v=0$ in the expression above and finding the respective $x_{max}$. He calls this point $c$  (see Fig. 5) but,
in a rather confusing way, he gives his answer in terms of the difference between the highest point in vacuum $AN$ and the highest point in the presence of air $An$.
So, following his line of thought, $x^{(vac)}_{max} - x^{(air)}_{max} = AN - An = Nn$ and he finally writes that there will be a point $c$ where the
pendulum will revert its motion. From this follows
\begin{equation}
\label{eq:anglearc}
 Nn = 2\frac{\gamma}{m}\times AN\times Ac + \frac{2\gamma\;a}{m}\times (nc - Ac)
\end{equation}
In corollary II Diderot gives an approximation for the above expression in terms of the results one would get in vacuum. He notes that the arc $Ac$ differs
very little from the vacuum value $AC$ and the same can be said about the height $nc$, which differs little from $NC$. So, he just rewrites the result above
replacing $nc$ by $NC$ and $Ac$ by $AC$ to get
\begin{equation}
 Nn = 2\frac{\gamma}{m}\times AN\times AC + \frac{2\gamma\;a}{m}\times (NC - AC)
\end{equation}
Corollary III consists in recasting the expression above when the oscillation amplitude is small, in which case $AC\approx NC$. This amounts to making
the last term on the right-hand side of the previous expression equal to zero and keeping only the first term~\footnote{In the original memoir there is
a misprint, when Diderot says that $AC$ should be `almost equal' to $AN$. One should substitute $AN$ by $NC$.}.

\begin{figure}[h]
\centering
\includegraphics[scale=0.5]{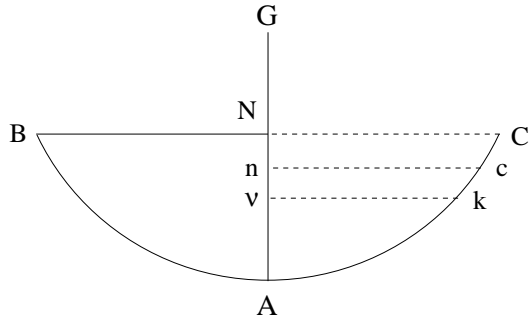}
\caption{The figure in Diderot's memoir depicting the highest point the bob reaches in the presence of air resistance. This is point $\nu$ on
the vertical axis, which corresponds to point $k$ of the trajectory.}
\label{diderotpendulumfig4}
\end{figure}

From these considerations Diderot now calculates the maximum height $A\nu$ the bob reaches when being let go from B (see Fig.~\ref{diderotpendulumfig4}).
Finding $\nu$ is equivalent to finding point $k$ along the trajectory to which it corresponds. With $k$ one may then calculate $Ck$, which is equivalent
to Newton's $RV$. To find this point Diderot uses the following argument:  
$C$ represents the point in the trajectory opposite to $B$, the starting position of the bob. As the bob is acted upon
by some drag, it will not reach $C$ but a certain $k$ below $C$. So, $Ck$ is the different in path between a bob with and without drag. However,
a bob falling from $B$ with air resistance is equivalent to a bob falling from an point below $B$
without air resistance. This point is the one opposite to $n$, between $C$ and $k$, in Fig. (\ref{diderotpendulumfig3}).

He now sums up his preceeding results: from Proposition I he found the velocity $v$ the bob has when reaching the lowest point $A$. It is the same
velocity the bob would have if it fell from a point below $B$ without air resistance. Calling now this velocity $h$, he uses it as a starting
velocity for the ascending bob.

So, from Cor. II of Prop. I one can say that
a bob falling from height $b=AN$ with air resistance is the same as falling from height $An<AN$ without air resistance:
\begin{equation}
\label{eq:aene}
 An = b- 2\frac{\gamma}{m}\times b\times BA - \frac{2\gamma}{m}\times a\times (BN - BA)
\end{equation}
Consequently, from Cor. II of Prop. II it follows that the bob will not go up to the opposite of point  because of air resistance,
but to a point $k$ (of height $A\nu$) slightly before $c$ (of height $An$).
\begin{equation}
\label{eq:ani}
 A\nu = An - 2\frac{\gamma}{m}\times An\times AC + \frac{2\gamma}{m}\times a \times (nc - Ac)
\end{equation}
Substituting in this expression the value of $An$ just found, and using the small angle condition such that $nc\approx BN$ and $Ac\approx BA$ one
ends up with
\begin{equation}
\label{eq:maximumheight}
 A\nu = b - \frac{4\gamma}{m}\times b\times BA+ \frac{4\gamma}{m}\times a \times (BN-BA)
\end{equation}
which is the content of Corollary III. Corollary IV is deduced from the fact that, when angles are small, $BN\approx BA$ and the
expression above reduces to
\begin{equation}
 \label{eq:maximumheight2}
 A\nu = b - \frac{4\gamma}{m}\times b\times BA
\end{equation}

%
%\subsection{Corollary V: The Devil in the Details }
The most important section of the Memoir, at least as to what regards Diderot's disagreement with Newton, is Corollary V.
He has to explicitly give an expression for $Ck$. From (\ref{eq:maximumheight}) one may get, for small angles ($BA\approx BN$) the simplified expression

\begin{equation}
 A\nu = b\biggl( 1- \frac{4\gamma}{m}\times BA\biggr) = AN\biggl( 1- \frac{4\gamma}{m}\times BA\biggr)
\end{equation}

\begin{figure}[h]
\centering
\includegraphics[scale=0.4]{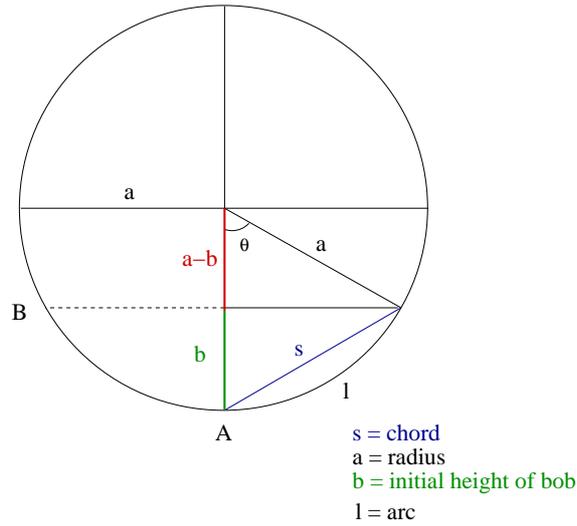}
\caption{The geometric construction to deduce Eq. (\ref{length}) of Diderot's memoir.}
\label{kreissehne}
\end{figure}

Now comes one section of Diderot's memoir which does justice to his style: what seems trivial is not worth explaining in more detail. He says
that for small angles, the arc $AC$ is to
$Ak$ as the root of $AN$ is to the root of $A\nu$. He adds: {\it ... since in the circle, the chords are among
them as the roots of the abscissae; or the arcs can be replaced here by the chords}.
%
%\begin{equation}
% AC = \sqrt{AN^2 + NC^2}\;\;\;\;\;\; and \;\;\;\;\;\; Ak = \sqrt{A\nu^2 + \nu k^2}
%\end{equation}
%It follows trivially that
%\begin{equation}
%\label{length1}
% \frac{Ck}{AC}=\frac{AC-Ak}{AC}=\frac{\sqrt{AN^2 + NC^2}-\sqrt{A\nu^2 + \nu k^2}}{\sqrt{AN^2 + NC^2}}
%\end{equation}
Diderot writes this as
\begin{equation}
\label{length}
 \frac{Ck}{AC}=\frac{\sqrt{AN}-\sqrt{A\nu}}{\sqrt{AN}}
\end{equation}
To see how one can get this, consider Fig. (\ref{kreissehne}).
The chord $s$ of Fig. (\ref{kreissehne}) can be written in terms of the radius $a$ and the angle $\theta$ by means of the cosine law
\begin{equation} 
s^2 = 2\;a^2 -2\;a^2\;\cos\theta
\end{equation}
Since $\cos\theta = (a-b)/a$ one may substitute this in the expression above to get
\begin{equation}
 s^2 = 2\;a^2 -2\;a^2\;\times\frac{a-b}{a}\longrightarrow s= \sqrt{2ab}
\end{equation}
Given that Diderot is considering small angles, one can approximate arcs by chords and thus write, in an approximate way
\begin{equation}
  l\sim \sqrt{2ab}
\end{equation}
From this one gets
\begin{equation}
 \frac{Ck}{AC}=\frac{AC-Ak}{AC}=\frac{\sqrt{2a\times\;AN}-\sqrt{2a\times\;A\nu}}{\sqrt{2a\times\;AN}}
\end{equation}
which is the same as Eq. (\ref{length}). By using the value of $A\nu$ found in Eq. (\ref{eq:maximumheight2}) and a series expansion for
the square root
\begin{equation}
\label{eq:squarerootapprox}
 \sqrt{1+x} \sim 1+ \frac{1}{2}x \;\;\;\;\;\; for \;\;\;\;\;\; x\ll 1
\end{equation}
Diderot arrives, after some straightforward algebra, at the result
\begin{equation}
\label{eq:diderotmainsolution}
 Ck=2\times\frac{\gamma}{m}\times (AB)^2
\end{equation}
which he expresses in terms not of $AC$, but of $AB$, since these arcs have the same length.
Thus difference in the arc due to air resistance is proportional to the square of the trajectory of the bob on its way down.
This is Diderot's main result and the point of the memoir where he confronts Newton.

If we compare his solution Eq. (\ref{eq:diderotmainsolution}) with Eq. (\ref{eq:diderotcompare}) obtained via a Lindstedt-Poincar\'e Method,
we can write the latter in Diderot's notation:

\begin{equation}
 Ck=\frac{4}{3}\times\frac{\gamma}{m}\times (AB)^2.
\end{equation}
where one can clearly see that prefactors of the two solutions differ. If we want to understand the reason why he did not get it
right one may look back at Eqs. (\ref{eq:anglearc}), (\ref{eq:aene}), (\ref{eq:ani}) and Fig.~(\ref{diderotpendulumfig3}).
The first of these is an equation for the point Diderot denotes by $n$ along
the vertical $OA$. This is the highest point the bob would reach had it started with velocity $h$ at the bottom. 
The expressions on both sides of this equation involve the unknown $n$, as it is hidden in the definition of the arcs
$nc$ and $Ac$. To make this point more clear, we can rewrite (\ref{eq:anglearc}) in terms of the variables $\theta_0$ (the angle of point $B$)
and $\theta_1$, the maximum value of $\theta$ on the bob's way up (to which height $n$ is associated). One obtains
\begin{equation}
 \cos\theta_1 -2\frac{\gamma}{m}a\,\sin\theta_1 +2\frac{\gamma}{m}a\,\theta_1= \cos\theta_0+2\frac{\gamma}{m}b\,\theta_0
\end{equation}
This is a transcendental equation for the unknown $\theta_1$. Instead of solving for $\theta_1$ (or $n$, which is the same),
he approximates $n$ by $N$ and $c$ by $C$ on the right-hand side of (\ref{eq:anglearc}). This is the same as
replacing $\theta_1$ by $\theta_0$ in those terms
\begin{equation}
\cos\theta_1 -2\frac{\gamma}{m}a\,\sin\theta_0 +2\frac{\gamma}{m}a\,\theta_0= \cos\theta_0+2\frac{\gamma}{m}b\,\theta_0 
\end{equation}
to get
\begin{equation}
\cos\theta_1= \cos\theta_0+2\frac{\gamma}{m}(b-a)\,\theta_0+2\frac{\gamma}{m}a
\end{equation}
He then subtracts
(\ref{eq:anglearc}) from $AN$ to get (\ref{eq:aene}). He then proceeds to (\ref{eq:ani}), keeping $n$ and $c$, but then
approximate them again to get (\ref{eq:maximumheight2}).
To conclude, Diderot's approach is to avoid solving the transcendental equation, by approximating his unknowns $c$ and $n$
by their values $C$ and $N$ in vacuum. By doing this he loses on the way  all important terms in the approximation which sum
up to give him the correct prefactor, while still getting the right functional dependence on $AB$.

There follows three colloraries (VI, VII and VIII)  which refer to ways of determining position 
$k$ (or $\nu$) in a back-of-the-envelope kind of calculation. They are straighforward consequences of the result he derived
in Corollary V. We reproduce them here for the sake of completeness.
\vskip 0.5cm\vskip 0.5cm
\noindent
\noindent
Corollary VI. If one knows the arc $ABC$ that a pendulum traverses when let go from $B$, one can easily find arc $bAk$ which is
the trajectory when let go from $b$. One just needs to find Ak, which one can get from
\begin{equation}
\frac{BA-AC}{bA-Ak}=\frac{BA^2}{bA^2}
\end{equation}
\vskip 0.5cm
\noindent
Corollary VII. Thus, if a pendulum falls through $BA$ in air, one can find its velocity in at point $A$ by dividing the $N\nu$
into two equal segments marked by point $n$. This is so since this velocity, according to Corollary III of Proposition I, is almost
the same as that obtained by a pendulum in vaccuum from point $b-(2\gamma/m)\times BA = b - N/2$.
\vskip 0.5cm
\noindent
Corollary VIII. One has
\begin{equation}
 \frac{AC^2}{Ac^2}=\frac{AN}{An}
\end{equation}
that is
\begin{equation}
 \frac{AC}{AC^2 -2Cc\times AC}=\frac{AN}{AN-Nn}
\end{equation}
from which follows
\begin{equation}
Nn= \frac{2Cc\times AC\times AN}{AC^2}=\frac{2Cc\times AN}{AC} 
\end{equation}
For the same reason one has
\begin{equation}
 N\nu= \frac{2Ck\times AN}{AC}
\end{equation}
and thus 
\begin{equation}
\frac{Ck}{Cc}=\frac{N\nu}{Nn} 
\end{equation}
Thus $c$ is the point in the middle of arc $Ck$. This means that, instead of dividing $N\nu$ into two equal parts, one may divide $Ck$ into
two equal parts in order to obtain the arc $Ac$ that body $A$ will have traversed in vacuum.

With these results he shows that if one consider a resistance force quadratic in the velocity, one indeed gets a retardation which is proportional
to the square of the arc $AB$. He further justifies his results with a bit dimensional analysis, before he moves on to his
{\it Eclaircissements}. His idea is the following: 
\begin{quote}
 {\it If pendulum $A$ is a small sphere, the resistance $f$, all other things being equal, is inversely proportional to the
 diameter of this sphere and its density; since the resistance caused by air on two spheres of different diameters goes
 as the surface or the square of the diameter; and this resistance has to be divided by the mass, that is like the density
 multiplied by the third power of the diameter. Thus the arc $Ck$, all other things being equal, is like $AB^2$ divided
 by the product of the diameter of the sphere and its density.}~\footnote{Si le pendule $A$ est un petit globe, la r\'esistance
 $f$, toutes choses d'ailleurs \'egales, es en raison inverse du diam\`etre de ce globe et de sa densit\'e; car la r\'esistance
 de l'air \`a deux globe de diff\'erents diam\`etres est comme le surface ou le carr\'e des diam\`etres; et cette r\'esistance
 doit \^etre divis\'ee par la masse, laquelle est comme la densit\'e multipli\'ee par le cube du diam\`etre. Donc l'arc $Ck$,
 toutes choses d'ailleurs \'egales, est come $AB^2$ divis\'e par le produit du dimat\`etre du globe et de sa densit\'e~\cite{diderot1}.}.
\end{quote}
How is this to be understood? Diderot is correct when he affirms that the resistance goes as the surface, as we now in hindsight that
it depends on the Reynolds number Eq. (\ref{eq:reynolds}). But when he affirms that `this resistance has to be divided by the
mass' it would mean, according to his reasoning, that
\begin{equation}
F_R\sim\frac{diameter^2}{mass}=\frac{diameter^2}{density\;\times\;diameter^3}= \frac{1}{density\;\times\;diameter}
\end{equation}
Diderot is not too rigorous with his wording, since from the sentence above the `resistance' $f$ cannot be the same $f$ he is using
to mean `resistance of air' throughout the text. He probably has as `acceleration' in mind. This is so since the drag force
depends only on the geometry of the bob. If they have the same diameters, the drag is the same. However, the equation of motion in the
two cases, given that they have different masses $m_1$ and $m_2$ is
\begin{equation}
 m_1 \; g\sin\theta - F_R = m_1\;a_{\theta}\;\;\;\;\;\;m_2 \; g\sin\theta - F_R = m_2\;a_{\theta}
\end{equation}
From which it trivially follows that the accelerations $a_{\theta}$ along the tangential of the arc are different in the two different
cases,
\begin{equation}
a_{1,{\theta}} = g\sin\theta - \frac{F_R}{m_1} \;\;\;and\;\;\;a_{2,{\theta}} = g\sin\theta - \frac{F_R}{m_2}
\end{equation}
Moreover, we know that the expansion of Lindstedt-Poincar\'e  is an approximation valid for small values
of $\epsilon=(\gamma\;l/m)$ which, for a fixed
string length $l$, takes exactly into account the ratio of the damping parameter and the bob's mass
(see discussion in Section~\ref{sec:MathematicalPendulum} below). His intuition got him on the right track.
\subsection{Diderot's \'Eclaircissements of Newton}

\begin{figure}[h]
\centering
\includegraphics[scale=0.4]{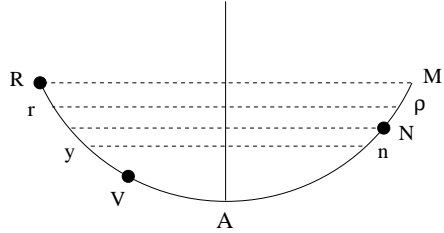}
\caption{The figure Diderot uses to explain his determination of the point $S$ in Newton's commentaries. In the figure in the original Memoir
there is a misprint: the letter $y$ should be opposite to $n$ and not $N$. The pendulum is let loose from $R$, reaches $N$ on the opposite side and return to $V$. For
the sake of clarity, these
points are marked here by black dots. $RV$ is the retardation of a full cycle.}
\label{pendulumdid5}
\end{figure}

As discussed in Section \ref{sec:NewtonSolution}, Newton explained the difference between experimental data and theoretical values in
his pendulum experiment as a consequence of air resistance. He gave an approximate value for $S$ (see Eq. \ref{eq:newtonvalueofs}) which
Diderot now calculates under the assumption of a linear drag. To better understand Diderot's solution, we reproduce Fig. \ref{pendulumdid5}
that Diderot uses in his Memoir while explaining the solution. 
\vskip 0.5cm
\noindent
{\bf Problem}: find the location of point $S$ such that a bob falling from it to point $A$ will have a retardation which is exactly equal to $1/4$
of a full cycle retardation $RV$.
\vskip 0.5cm
\noindent
In his Memoir Diderot chooses the arcs such that $RA = 1$, $RV=4b$, and $SA=x$. He sets out to find $x$. The choice of variables is an indication that
he also studied the {\it Principia} from the annotated editions of Le Seur and Jacquier. They say, when referring to this passage of Newton
that Diderot is addressing:
\begin{quote}
{\it Bring body $A$ to any point $R$ along the arc $EAF$ and let it fall
from there. If the resistance of the medium is absent, it will reach
the same height $M$ to which it was lifted and should return to $R$.
But when, after the first oscillation composed of exit and  return,
it returns to point $V$ (according to the hypothesis), the arc $RV$ will represent the retardation 
of a double ascent and descent [caused by the] medium; thus one should take the retardation due to
the medium in one whole descent as the fourth
part of the total retardation, that is the fourth part of arc $RV$, provided it
did not descend neither from the highest point $R$ nor from the lowest $V$ to begin
with: for the retardation will be larger for the larger arc than the smaller one, since as the
pendulum describes ever smaller oscillations, the retardation of each single arc will be unequal,
and the retardation of the descent by $RA$ will be bigger than the fourth part of $RV$, and
the retardation of the last ascent $AV$ will be smaller than the fourth part of the total
retardation $RV$. With a similar calculation Newton determined a point $S$ such that the
retardation in descending through $SA$ should be [exactly] the fourth part of of the total
retardation $RV$. Let arc $RA$ be $1$, arc $RV$ be $4b$ and the arc sought $SA$ be $x$;
since the retardation is proportional to the arc [traversed], the arc $SA$ ($x$) is to the arc $RA$ ($1$),
as the retardation of the arc $SA$, defined as $b$, the fourth part of the whole $RV$, 
is to the retardation of the first arc $RA$, that is $b/x$. 
The successive delays to be found, the second, the third and the fourth follow
the same ratio. The arc of the second is equal to $RA$, allowing for
a retardation of $b/x$. The third arc is equal to the second, allowing for
the same retardation, and so on, but all these delays sum up to give the whole $RV$, or $4b$.
Hence [we obtain] the equation from which we get the value of the arc $SA$, or $x$, [which] by means of an
approximation [turns out] to be equal to $1$ $3/2\,b$ [that is $RA$ $3/2\,b$].
So taking $RS$ equal to the fourth part of the arc
$RV$ with its half, the retardation of the arc 
$SA$ will be  equal to $ST$, the fourth part of the total retardation $RV$, and therefore 
a body dropped from point $S$ will have the same speed at $A$, without significant error,
as it would have if it had fallen in vacuum from $T$.
~\footnote{\it Trahatur corpus $A$, ad arcus $EAF$,
punctum quodvis $R$, et demittatur inde, sublata
medii resistentia ad eandem altitudinem $M$, ascendere
et rursus ad punctum $R$, redire debet.
Cum autem post unam oscillationem exitu et reditu compositam perveniat (ex hyp.)
ad punctum $V$
arcus $RV$ exponet medii retardationem
in duplici ascensu et descensu; quare ut
habeatur medii retardatio in uno tantum descensu,
sumenda est quarta pars totius retardationis,
id est quarta pars arcus $RV$, dummodo
ille descensus neque ex puncto supremo $R$, neque
ex infimo $V$ ordiatur: nam cum major sit
medii retardatio in arcu majori quam in minori
semperque fiant minores arcus a pendulo oscil
lante descripti, inaequales quoque erunt retarda
tiones in singulis arcubus, et retardatio descensus
per $RA$, major erit quarta parte totius
retardationis $RV$ ut retardatio ultimi ascensus
$AV$, minor erit quarta parte totius retardationis
$RV$. Hoc autem aut simili calculo determinavit
Newtonus punctum $S$ tale ut retardatio in
descensu per $SA$ sit quarta pars totius retardationis
$RV$. Dicatur arcus $RA$, $1$, arcus $RV$,
$4 b$, arcus quaesitus $SA$, $x$; sintque retardationes
arcubus
descriptis proportionales, erit arcus $SA$
($x$) ad arcum $RA$ ($1$) ut retardatio arcus $SA$
quae statuitur esse $b$, seu quarta pars totius
$RV$, ad retardationem primi arcus $RA$ quae erit
$b:X$. Quaerantur successive retardationes secundi, 
quartive arcus eadem ratione ; arcus autem 
secundus est equalis primo $RA$, dempta ejus retar- 
dationes $b:X$. Tertius arcus aequalis secundo demp- 
ta ejus retardatione, et sic deinceps, omnes vero 
illae retardationes simul sumptae aequabuntur toti 
retardationi $RV$ seu $4 b$ ; unde fit aequatio ex qua 
valor arcus $S A$, seu $x$, obtinebitur, per approxi-
mationem autem invenietur aequalis $1\; 3/2 b$
sumatur itaque $RS$ aequalis quartae parti cum ejus
semisse totius retardationis $RV$, retardatio per
arcum $SA$ erit aequalis $ST$ quartae parti totius
$RV$, ideoque cadat corpus ex puncto $S$, ejus celeritas in $A$ eadem est sine errore 
sensibili, ac si in vacuo decidisset ex $T$~\cite{jacsu}.}}
\end{quote}
 
This is the origin of Diderot's naming of arcs and is more explicit than Newton in indicating how the calculation should be done.
It does not spare him however the work of actually finding $S$. 
For the sake of completeness and further comparison of lengths, we will keep $RA$ arbitrary while maintaining Diderot's value for $RV$ and $SA$, that is $4b$
and $x$. We closely follow Diderot's ideas up to his solution.

If a body falls from $A$, it will return to $V$. Each quarter cycle contributes a retardation $r_i$ such that
\begin{equation}
 r_1+r_2+r_3+r_4= RV
\end{equation}
We know that the retardations are not equal, that is $r_i\ne r_j$ for $i \ne j$. As an approximation, we can think of a point below
$R$ -- call this point $S$ -- such that
a bob falling from it until $A$ will show a quarter-cycle retardation exactly equal to $(1/4) RV = b$. This must be so since we know that a body falling from
$R$ will have a retardation $r_1^{(R)} > b$, so to have a retardation smaller requires $S$ to be further down the track. Newton's hypothesis on the 
retardation is that is proportional to $RA$, 
\begin{equation}
\label{eq:linearassumption}
 Rr = \alpha RA
\end{equation}
where $\alpha$ is some constant. It is important to recall that Newton's argument is based on the idea that a bob falling with resistance from $R$ is the
same as falling without resistance from a lower point $r$. So if one determines the velocity $v_A$ at $A$ in air, one can just use some reverse engineering
and determine which $r$ would give that same $v_A$ in vacuum. This is trivial, since then one may just use conservation of mechanical energy to find $r$.
Following this idea, Diderot assumes that a bob falling from $S$ would be the same as a bob falling in vacuum from $r'$ and, given Newton's assumption,
one would have in place of the equation above
\begin{equation}
 Sr' = \alpha SA
\end{equation}
where $Sr'$ is the retardation when falling from $S$. But the problem is to find $S$ for which this retardation is exactly $b$. So, by eliminating $\alpha$
in the equations above one gets
\begin{equation}
 \frac{Rr}{RA}=\frac{Sr'}{SA}\longrightarrow Rr=\frac{Sr'}{SA}\;RA
\end{equation}
Since $Sr'=b$ and $SA=x$ this reduces to
\begin{equation}
 Rr= RA - rA = \frac{b}{x}\;RA
\end{equation}
Now, the arc described on the first ascent would be $A\rho = Ar = (1-b/x)\,RA$ but due to air resistance the bob does not reach $\rho$ but a lower point $N$
such that
\begin{equation}
 \rho N = \alpha A\rho =\alpha Ar=\alpha\biggl( 1-\frac{b}{x}\biggr)\;RA =\frac{b}{x}\biggl( 1-\frac{b}{x}\biggr)\;RA
\end{equation}
So the actual arc the bob describes is
\begin{equation}
 AN = A\rho -\rho N = \biggl( 1-\frac{b}{x}\biggr)\;RA -\frac{b}{x}\biggl( 1-\frac{b}{x}\biggr)\;RA=\biggl( 1-\frac{b}{x}\biggr)^2\;RA
\end{equation}
So, by following this kind of argument one can determine all four quarter cycle retardations. They are
\begin{eqnarray}
 Rr &=& \frac{b}{x}RA\nonumber\\
 \rho N &=&\frac{b}{x}\biggl( 1-\frac{b}{x}\biggr)\;RA\nonumber\\
 Nn &=&\frac{b}{x}\biggl( 1-\frac{b}{x}\biggr)^2\;RA\nonumber \\
 Vy &=& \frac{b}{x}\biggl( 1-\frac{b}{x}\biggr)^3\;RA
\end{eqnarray}
The sum of all these retardations should be $4b$, that is
\begin{equation}
\label{eq:prequartic}
 RA\biggr[\frac{b}{x} +\frac{b}{x}\biggl( 1-\frac{b}{x}\biggr)+\frac{b}{x}\biggl( 1-\frac{b}{x}\biggr)^3
+\frac{b}{x}\biggl( 1-\frac{b}{x}\biggr)^3\biggr] = 4b
\end{equation}
This leads to a quartic equation in the unknown $x$
\begin{equation}
\label{eq:quartic}
 \frac{1}{RA}\,x^4-x^3 + \frac{3b}{2}x^2 -b^2 x +\frac{b^3}{4}=0
\end{equation}
Before solving this equation, Diderot considers the limiting case where $b\ll 1$, in which case one may neglect the last two terms
on the right hand side and write
\begin{equation}
 \frac{1}{RA}\,x^4-x^3 + \frac{3b}{2}x^2=0\longrightarrow x^2 - RA\,x +\frac{3b}{2}\,RA=0
\end{equation}
This equation has two solutions, namely
\begin{equation}
 x_{+,-}= \frac{RA}{2} \pm \frac{RA}{2}\sqrt{1-\frac{6b}{RA}}
\end{equation}
If one further considers an approximation to the square root given by Eq. (\ref{eq:squarerootapprox}) these solutions reduce to
\begin{eqnarray}
 x_+ &=& RA -\frac{3}{2}b\nonumber\\
 x_{-}&=& \frac{3}{2}b
\end{eqnarray}
Solution $x_{-}$ is not physically acceptable, since it would imply that $S$ is close to $A$. Solution $x_{+}$ can be written as
\begin{equation}
\label{eq:approximatex}
 x_+ = RA -\biggl(b + \frac{1}{2}b\biggr)
\end{equation}
If one recall that $b$ is what Newton called $(1/4) RV$, one can write
\begin{equation}
 x_+ = RA -\biggl(\frac{1}{4}RV + \frac{1}{8}RV\biggr)
\end{equation}
which is the same as Eq. (\ref{eq:newtonvalueofs}). So, Diderot shows that Newton's placement of $S$ can be recovered in the limit where
$b$ is taken as being very small. But Diderot goes a bit further, by solving exactly the quartic. He does this by expanding the exponents
in (\ref{eq:prequartic}) and rewriting it as
\begin{equation}
\biggl(4\frac{b}{x} -6\frac{b^2}{x^2}+4\frac{b^3}{x^3}-6\frac{b^4}{x^4}\biggr)= \frac{4b}{RA}
\end{equation}
He then notices that 
\begin{equation}
1-\biggl(4\frac{b}{x} -6\frac{b^2}{x^2}+4\frac{b^3}{x^3}-6\frac{b^4}{x^4}\biggr)= \biggl( 1-\frac{b}{x}\biggr)^4
\end{equation}
and therefore
\begin{equation}
 \biggl( 1-\frac{b}{x}\biggr)^4 = 1- \frac{4b}{RA}
\end{equation}
This equation has four solutions. Two are pure imaginary and can be discarded. From the two real solutions the one which is
physically relevant is
\begin{equation}
\label{eq:exactsolutionofx}
 x= \frac{b}{1-\biggl(1-\frac{4b}{RA}\biggr)^{\frac{1}{4}}}
\end{equation}
This is the exact position of point $S$.

It is important to note that Diderot does not use his previous results (Propositions I and II) in order to obtain the position of point
$S$. The fact is that he does not need to: Newton's result follows from the simple assumption that retardation is proportional to the arc.
But since Diderot wrote also his article arguing for a retardation proportional to the square of the arc, that is
\begin{equation}
 Rr = \alpha (RA)^2,
\end{equation}
why didn't he bother to write down the equation that would replace (\ref{eq:quartic}) and solved it? The new equation can be written in a
straighforward manner, albeit after a very long algebraic manipulation. One obtains
\begin{eqnarray}
 4bx^{30} - 4 a^{2}bx^{28} + 12\;a^{3}b^{2}x^{26} - 30\;a^{4}b^{3}x^{24} + 64\;a^{5}b^{4}x^{22} - 118\;a^{6}b^{5}x^{20}&+&\nonumber\\
 188\;a^{7}b^{6}x^{18} - 258\;a^{8}b^{7}x^{16} + 302\; a^{9}b^{8}x^{14} - 298\;a^{10}b^{9}x^{12} + 244\;a^{11}b^{10}x^{10}&-&\nonumber\\
 162\; a^{12}b^{11}x^{8} + 84\;a^{13}b^{12}x^{6} - 32\;a^{14}b^{13}x^{4}  + 8\;a^{15}b^{15}x^{2} - a^{16}b^{15} &=&0
\end{eqnarray}
where, for the sake of clarity, we replaced $RA$ by the letter $a$. We don't know if Diderot ever wrote this equation but in any case it
does not appear in the memoir. This is not surprising and it is quite pointless trying to find the roots to this equation. We can however try to find an approximate solution
of the reduced polynomial: if we, with Diderot, consider $b$ to be small, we can discard the higher powers of $b$ and keep only the terms
to lowest order, that is
\begin{equation}
 4b\; x^{30} - 4 a^{2}b\;x^{28} + 12\;a^{3}b^{2}\;x^{26} =0 \longrightarrow x^4 - a^{2}x^2 + 3\;a^{3}b =0
\end{equation}
This equation can be trivially solved to give the roots
\begin{equation}
 x=\pm \frac{a}{\sqrt{2}}\sqrt{1\pm\sqrt{1-\frac{12b}{a}}}
\end{equation}
Again, approximating the square root as in Eq. (\ref{eq:squarerootapprox}) one gets a physically relevant solution in the form
\begin{equation}
 x = a - \frac{3}{2}b
\end{equation}
Remembering that $a$ is our short notation for $AR$ this solution can be written as
\begin{equation}
\label{eq:approximatexdid}
x = RA -\biggl(b + \frac{1}{2}b\biggr)
\end{equation}
which is the same approximate solution Eq. (\ref{eq:approximatex}] that Newton got in the case of a retardation proportional to the arc. To conclude,
in the limit of small amplitude oscillations, where velocities are small,
replacing a linear by a quadratic drag makes no significant difference. This is what one observes in the experiments discussed
in the last section of this article: the changes in period due to linear and quadratic drag are of the same order of magnitude
and it would have been impossible for Diderot or Newton to detect those.

Another interesting point worth noticing is the fact that Diderot did not apply the method he develops 
for the case of a linear drag. He must have been aware that this would imply replacing the integral
in Eq. (\ref{eq:diderotintegral}) by
\begin{equation}
\int_{b}^{x}\frac{\sqrt{2pb-2px'}\;a\;dx'}{\sqrt{2ax' - x'^2}},
\end{equation}
which can be solved only numerically~\footnote{It can be recast in terms of a very complicated expression involving an elliptic integral of the second kind,
whose values can then be looked up in a table or solved numerically. Elliptic integrals go back to A.-M. Legendre's (1853 -- 1833) and N.H. Abel's (1802 -- 1829)
works of 1825 and 1823, respectively~\cite{good}.}.
%
%=================================================================================
\section{The Mathematical Pendulum from a Modern Perspective}
\label{sec:MathematicalPendulum}
%==========================================================================
\subsection{The Problem and the Solution}
The mathematical (ideal) pendulum is one of the most paradigmatical models of classical mechanics. It consists
of a pointlike mass $m$ attached to a frictionless point $O$ through an ideal (massless and inextensible) string of length $l$.
As it swings, the position of the bob can be described, for
any given instant $t$, by the angle $\theta\;(t)$ measured relative to its rest position $A$ (see Fig. \ref{mathematicalpendulum}). 
\begin{figure}[h]
  \centering
  \includegraphics[scale=0.5]{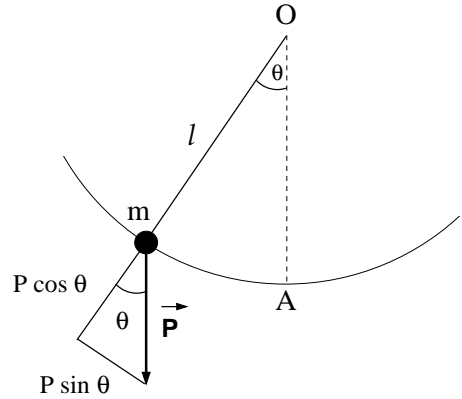}
  \caption{The ideal or mathematical pendulum. A pointlike mass $m$ attached to a point O through an ideal string of length $l$. $\theta$ is the angle the bobs
  makes with respect to the vertical $OA$. The angle $\theta$ is positive if to the right of the vertical $OA$ and negative to the left.}
\end{figure}
\label{mathematicalpendulum}
\noindent
Even though one can write Newton's equation of motion for the displacement $d\vec{s}$\, along the arc, it is more convenient to write
the same equation in terms of the angle $\theta$ ($ds = l\;d\theta$)
\begin{equation}
\label{eq:fullpendulum}
 m\;l\;\frac{d^2\theta}{dt^2} + m\;g\;\sin\theta  = 0 
\end{equation}
This nonlinear differential equation has an implicit solution for $\theta$ as a function of time $t$ in terms of
Legendre's elliptic integral of the first kind $F(k,\psi)$~\cite{fluegge}
\begin{equation}
 \label{eq:solutiongeneral}
\sqrt{\frac{g}{l}}\;t = \int_{0}^{\psi}\frac{d\psi\;'}{\sqrt{1-k^2\sin^2\psi\;'\;^2}} = F(k,\psi)
\end{equation}
The angular displacement $\theta$ is related to  $\psi$ through
\begin{equation}
\label{eq:thetaalphapsi}
 \sin\frac{\theta}{2} = k\sin\psi ,
\end{equation}
where $k$ is a quantity related to the amplitude $\theta_0$ through
\begin{equation}
k= \sin\frac{\theta_0}{2}
\end{equation}
As the bob is let loose from $\theta_0$, it will swing and by solving the above integral numerically, one can determine
the value of $\theta (t)$ at any given time $t$. One is normally interested in the period $T$ of one complete oscillation.
From the result above one may easily obtain
\begin{equation}
\label{eq:periodgeneral}
 T = 4\sqrt{\frac{l}{g}}\;\int_{0}^{\pi/2}\frac{d\psi\;'}{\sqrt{1-k^2 \sin^2 \psi\;'^2}} = 4\sqrt{\frac{l}{g}}\;K(k)
\end{equation}
where $K(k) = F(k,\pi/2)$ is the complete elliptic integral of the first kind. This result follows by remembering that
a period corresponds to the time it takes the bob to return to $\theta_0$ after its release. In this case $\theta(T) = \theta_0$
implies $\psi = \pi/2$ in Eq. (\ref{eq:thetaalphapsi}) and hence $F(k,\psi)\rightarrow  F(k,\pi/2) = K(k)$.

This solution is rather involved and what one will usually find in physics textbooks is the small amplitude approximation: in the case of small $\theta$,
one may replace $\sin\theta\approx\theta$ in Eq. (\ref{eq:fullpendulum}) and obtain a {\it linear} differential equation
\begin{equation}
\label{eq:simplifiedpendulum}
\frac{d^2\theta}{dt^2} +\frac{g}{l}\theta  = 0 
\end{equation}
which can be easily solved to give (for the initial condition $\theta (t=0) = \theta_0$)
\begin{equation}
\theta (t) = \theta_0\cos\biggl(\sqrt{\frac{g}{l}}\;t\biggr)
\end{equation}
The period $T$ is then given by
\begin{equation}
\label{eq:period}
 T = 2\pi\sqrt{\frac{l}{g}}
\end{equation}
The relevance of this exact solution in the small-$\theta$ limit lies not only in its use as a didactic tool in the study of differential equations.
It shows that for small angles, the period of the pendulum Eq. (\ref{eq:period}) depends only on the length $l$ of the string and the acceleration of gravity $g$
and not on the amplitude of the swing. This makes the small-amplitude pendulum
an ideal time-keeping device.% Newton also used this result as evidence of the proportionality between gravitational and inertial masses
%~\footnote{If one considers Newton's equation of motion in the small amplitude approximation considering that these two masses might be different

%\begin{equation}
%\label{eq:simplifiedpendulum2}
%m_{G}\frac{d^2\theta}{dt^2} + m_{I}\;(g/l)\;\theta = 0 
%\end{equation}

%where $m_{I,G}$ stands for inertial (gravitational) mass, one obtains a period $T$ of the form

%\begin{equation}
%\label{eq:period2}
% T = 2\pi\sqrt{\frac{m_{G}}{m_{I}}\frac{l}{g}}
%\end{equation}

%Once $l$ and $g$ are known, this equation can be used to test whether $m_I = m_G$.}

By measuring the period and the length of the pendulum,
one may also use eq. (\ref{eq:period}) to find the accelaration of gravity $g$ with quite good precision, and this method
was the preferred one before being substituted by direct measurements on free-falling bodies~\cite{no}.

The small-amplitude approximation can be obtained from the general solution Eq. (\ref{eq:periodgeneral})  by rewriting it as a power series in
$k= \sin(\theta/{2})$
\begin{equation}
\label{eq:periodexpansion}
 T = 2\pi\sqrt{\frac{l}{g}}\biggl(1+\frac{1}{4}\;k^2+\frac{9}{64}k^4 + \dots \biggr)
\end{equation}

For an amplitude of $\theta_0 = 20$\,\textdegree\, ($k=0.1736$) the correction that Eq. (\ref{eq:periodexpansion}) introduces
amounts to about $0.7$\% as compared to
Eq. (\ref{eq:period}) while for $\theta_0 = 40$\,\textdegree\, ($k=0.3420$) it amounts to about $3$\%. At the time Diderot wrote his memoirs
the solution of the pendulum equation for arbitrary angles was not known.
%
%=================================================================== 
\subsection{The Effect of Air Resistance}
%===================================================================
% 
The simplicity of the approximate solution is quite deceptive not because the small angle approximation is unphysical -- for this one may always
go back to the entire solution - but because
in real applications the ideal conditions assumed from the onset are not valid: bobs are not pointlike masses, strings are not massless and inextensible and
damping by air and friction at the pivoting point do play a role. The bob will eventually stop swinging if there is no external force to keep it moving.
%From all possible sources of correction to the movement of the pendulum%, extensively discussed by Nelson and Olsson~\cite{no},
%the one most directly concerned with Diderot's work are the so-called air corrections, which can be basically classified in three types:
%\begin{itemize}
% \item[(i)] Buoyancy: Archimedes's principle teaches us that the volume of air displaced by the bob generates a buoyancy force opposite to the bob's weight,
% lengthening the period. For a normal pendulum clock this would mean a loss of the order of $7\;s$ a day~\cite{no}. 
 %
% \item[(ii)] `Added mass effect': part of the kinetic energy of the pendulum's bob is imparted to the
% surrounding air. By conservation of energy, the kinetic energy thus acquired by the bob will be smaller and this can be modelled by considering that the bob
% has extra mass added to it. This correction was first noted by Bessel in 1828~\cite{bessel} and before him by Du Buat in 1786, two years after Diderot's death.
% This effect amounts to a clock deviation of the order of $8\;s$ in a day~\cite{no}
 %
% \item[(iii)] 
 For the present work, the most relevant source of damping is the resistance caused by the surrounding air. Diderot, as did Newton before him,
  considered the effect of air resistance on a spherical bob, in spite of the fact that it also acts on the wire from which the bob hangs (see discussion below).
  The main question that puzzled physicists for a long time and can be only effectively dealt in a phenomenological manner is how the drag (as the force
  due to air resistance is usually called in technical parlance) depends on the relative speed between the moving body and the surrouding air. This is the main
  point of divergence between Diderot and Newton and the correct answer to this question is not or mere academic or historical interest: it  has consequences
  that go beyond the problem discussed here, as for instance in the design of aircraft wings or of any object that moves through air. 
%\end{itemize}

We know today a lot more about the effects of drag than Diderot (or for that matter Newton) knew at the time he wrote his memoir. 
Hydrodynamics was still on the making and ideas and techniques which allowed one to handle the effect of air resistance was developed mostly during the 18th and 19th centuries by people like G. G. Stokes (1819 -- 1903),
J. W. Struth (Lord Rayleigh, 1842 -- 1919), O. Reynolds (1842 -- 1912) and L. Prandtl (1875 -- 1953). What determines the type of drag acting on any part
of a system is determined by the Reynolds number characteristic of that part (for the pendulum, these would be  the bob and the string).
This dimensionless quantity was first introduced by Stokes in 1858~\cite{stokes} to predict flow patterns in fluids but was named after Reynolds, who popularized
its use in 1883~\cite{reynoldscite}. The Reynolds number is defined as
\begin{equation}
\label{eq:reynolds}
Re := \frac{\rho v L}{\eta}
\end{equation}
where $\rho$ and $\eta$ represent the density and dynamic viscosity of the surrouding medium respectively and $v$ is a characteristic velocity, usually the mean
relative velocity between fluid and body. $L$ is a characteristic length of the body, which in the case of the bob would be its diameter.

The Reynolds number is the key as to whether the drag be Stokes-like or Newton-like.
Low Reynolds numbers, up to $Re\sim 10$, imply that the flow past the body will be laminar (not turbulent) and will cause a drag
of the Stokes type. This is the case for instance of a ball falling through honey (low velocity, high viscosity), where it quickly reaches a terminal velocity and
then falls at constant speed. High values of $Re$, of order $\sim 10^3$ to $10^5$ (high velocity, low viscosity) corresponds to a drag force which follows Newton's
$v^2$--law. As the pendulum consists of two parts, a string and a bob, in a real experimental setup one has to treat each component according to its own Reynolds
number. As an example, in the experiments conducted by Nelson and Olsson, the string had a  Reynolds numbers of $6$ while the bob had an $Re$ of $1100$. This implies that the best fit to the problem of a swinging bob
would be~\cite{no}
\begin{equation}
\label{eq:drag1}
F(v) = a\;|v| + b\; v^2
\end{equation}
where $a,b$ are adjustable parameters and the first term on the right-hand side accounts for the drag on the string while the second for the drag on the bob. For the
experiment conducted by Nelson and Olsson, $a$ is usually one order of magnitude smaller than $b$, so by choosing a string thin enough, one would not be too far off the mark if one
just considered $a=0$ and took $F(v) = b\; v^2$, as Diderot did.

Another point that makes matters significantly more difficult as was not explicitly discussed in the previous literature on Diderot is the fact that,
as the bob swings, its velocity changes and so does its Reynolds number. The usual heuristic approach to deal with this problem is to consider a
generalization of Eq. (\ref{eq:drag1}) in the form
\begin{equation}
\label{eq:drag2}
 F = \frac{1}{2}\; C_D\;A\rho\;v^2
\end{equation}
where the dimensionless number $C_D$, known as the drag coefficient, incorporates the effect of a changing Reynolds number. In the expression above
$A$ is an area associated with the moving body. $C_D$ is a function of the Reynolds number
and is determined by adjusting experimental data for $F$ as a function of $v$.  For values of $Re$ of the order of 1 or smaller, $C_D$ is
inversely proportional to $Re$, that is $C_D\sim Re^{-1}$ while for high values of $Re$, $C_D$ is a constant.
This way one tries to capture the whole range of regimes under one single equation. For the case of a spherical bob, an expression for $C_D$ accurate
to within $10$\% for values or $Re$ over the range $0\leq Re \leq 2\times 10^5$ can be found in~\cite{white} and is given by
\begin{equation}
 C_D\simeq \frac{24}{Re} + \frac{6}{1+Re^{1/2}} +0.4
\end{equation}
where the first term on the right-hand side accounts for Stokes's law, the last for Newton's $v^2$ law and the middle term for the transition between
both regimes. 

So the question about a $v$ or $v^2$--dependence is not straighforward, as already
pointed out by~\cite{kk} and~\cite{coolidge}. However, by assuming an drag of the type given by
Eq. (\ref{eq:drag2}), Diderot sounds surprisingly modern. What about Newton? Even
though Newton he does not refer to the size of the bob he used in trying to prove the Third Law, in Book II he is explicit about the size of
pendulums he experimented with: a wooden bob of approximate diameter of $d=17.46\;cm$ and mass $m= 1625\;g$ and a leaden bob of
$d=5.08\;cm$ and $m= 744.2\;g$. This would imply that Newton's bobs had a range of $Re$ from approximately $1100$ to $3700$, which calls
for a $v^2$--law, assuming that he used the same bob sizes in his collision experiments. Coincidentally his second bob is
about the same size and mass as the one used by Nelson and Olsson in their experiments, so we can use their results to see how far off Diderot
or Newton might have been~\cite{no}.

What the experimental results show is that the corrections are of the same order of magnitude, irrespective of whether
one considers the first type of force (Newton) or the second (Diderot). In their experiments, Nelson and Olsson took an initial amplitude of $3${\textdegree}$\pm 0.3${\textdegree}
which introduces a finite-amplitude correction of $596\;\mu s$ when compared to the ideal period. In the case of linear damping the correction to the period,
discounting the finite-amplitude correction, was of the order of $0.033\;\mu s$. For the quadratic and one it amounted to $0.027\;\mu s$. The difference is
negligible and it would have been impossible for Diderot (or Newton) to detect those~\footnote{In the experiment one does not measure the effect directly. One
considers a drag of the type $F(v) = a\;|v| + b\; v^2$  and finds the best values of $a$ and $b$ that fit the data. From that one can inferr the retardation
effect using the approximate solutions with the fitted values. There is an added complication, since as the pendulum swings and is damped,
the amplitude changes and consequently the period. To measure the effect of air damping one has to average over many oscillations and discount the finite
amplitude correction accumulated during swings.}. This is reflected in the equality of Diderot's solution for Newton's $S$, Eq. (\ref{eq:approximatex})
in the linear case and our solution Eq. (\ref{eq:approximatexdid}) in the quadratic case.

\subsection{The Small-Angle Approximation and Quadratic Damping: the Method of Lindstedt-Poincar\'e}

Unbeknown to Diderot, he was trying his hand at a problem whose exact solution still eludes us.
If one looks up any textbook on the effect of air resistance on a pendulum -- and in most textbooks the pendulum equation
means the linearized version (\ref{eq:simplifiedpendulum}) and not Diderot's nonlinear Eq. ({\ref{eq:fullpendulum}) -- one will always find a
{\it linear} drag force $F_R (v)$ and not a quadratic one. To the author's knowledge, none of the texts consulted give any justification,
experimental or otherwise, as to why this should be so. The reason might be purely didactical: if one considers a drag force
$F_R (v)= -\gamma v$, Eq. (\ref{eq:simplifiedpendulum}) becomes
\begin{equation}
\frac{d^2\theta}{dt^2} + \frac{\gamma}{m}\;\frac{d\theta}{dt} + \frac{g}{l}\;\theta  = 0 
\end{equation}
for which one may easily find an exact analytical solution with an exponentially damped amplitude~\cite{fluegge}.
On the other hand, if one considers quadratic damping
\begin{equation}
\label{eq:pendulumquadratic}
\frac{d^2\theta}{dt^2} - \frac{\gamma\;l}{m}\biggl(\frac{d\theta}{dt}\biggr)^2 + \frac{g}{l}\;\theta  = 0 
\end{equation}
there is no exact solution anymore. What is worse, the equation is not even analytic because the sign of the force (and hence the equation) must be adjusted each
half-period to guarantee that the damping force always acts as to retard the pendulum's movement.
It took a century after Diderot's death for A. Lindstedt (1854 -- 1939) and H. Poincar\'e (1854 -- 1912)to independently develop a method that allows
one to treat the problem in a perturbative way. As the result so obtained is important to understand Diderot's solution. We closely follow the solution
as presented in ~\cite{no,linz,cve}.

The difficulty with Eq. (\ref{eq:pendulumquadratic}) is that standard perturbation methods will not work. This is because there are two time scales involved,
the one associated with the period of the pendulum and the other with dissipation. A standard perturbation method leads to the appearance of so-called
secular terms, which are terms which grow with time, whereas one knows that the solution has to be periodic. The Lindstedt-Poincar\'e method is a way of
removing these secular terms when dealing with weakly nonlinear problems with periodic solutions.

We consider Eq. (\ref{eq:pendulumquadratic}) for a half-period of oscillation, since the solution obtained can be reapplied to other half-periods. We
rewrite this equation as
\begin{equation}
\label{eq:pendulumquadratic2}
\ddot{\theta} - \epsilon\;\dot{\theta}^2  + \omega_{0}^2\;\theta  = 0 
\end{equation}
where $\dot{\theta}=d\theta/dt$, $\epsilon=(\gamma\;l/m)$ and $\omega_{0}^2 = g/l$. We want to find a solution with a period $T=2\pi/\omega$. One introduces
a new variable
\begin{equation}
 \phi = \omega t
\end{equation}
in terms of which Eq. (\ref{eq:pendulumquadratic2}) can be written as
\begin{equation}
\label{eq:pendulumquadratic3}
\omega^2\theta'' - \epsilon\omega^2{\theta '}^2  + \omega_{0}^2\;\theta  = 0 
\end{equation}
where now $\theta '$ stands for $d\theta/d\phi$. The following step is to write $\theta$ and $\omega$ in terms of a series expansion in
the small parameter $\epsilon$
\begin{eqnarray}
\label{eq:seriesexpansion}
 \theta&=&\psi_0+\epsilon\psi_1+\epsilon^2\psi_2+\cdots\nonumber\\
 \omega&=&\omega_0+\epsilon\omega_1+\epsilon^2\omega_2+\cdots
\end{eqnarray}
and then substitute (\ref{eq:seriesexpansion}) into (\ref{eq:pendulumquadratic3}). Setting the factors of each power of $\epsilon$ equal to zero, we
obtain, through order $\epsilon^2$, the following set of equations
\begin{eqnarray}
\label{eq:seriesexpansion2}
 \psi_0''+\psi_0 &=& 0 \nonumber\\
 2\biggl( \frac{\omega_1}{\omega_0}\biggr) \psi_0'' + \psi_1''+\psi_1-\psi_0'\,^2 &=& 0 \nonumber\\
 \biggl[ 2\biggl(\frac{\omega_2}{\omega_0}\biggr) + \biggl(\frac{\omega_1}{\omega_0}\biggr)^2 \biggr]\psi_0'' +
 2\biggl( \frac{\omega_1}{\omega_0}\biggr) \psi_1''&+& \nonumber\\
\psi_2''+\psi_2 -2\biggl( \frac{\omega_1}{\omega_0}\biggr) \psi_0'\,^2 -2\psi_0'\,\psi_1'&=& 0 
\end{eqnarray} 
These can be solved recursively. The solution of the first equation with $\psi_0=\theta_0$ and $\psi_0'=0$ at $\phi = \omega t = 0$ is
\begin{equation}
 \psi_0 = \theta_0\cos\phi
\end{equation}
Substituting this into the second equation in (\ref{eq:seriesexpansion2}) leads to
\begin{equation}
\label{eq:psiequation}
 \psi_1''+\psi_1= 2\biggl( \frac{\omega_1}{\omega_0}\biggr)\theta_0 \cos\phi+\frac{1}{2}\theta_0^2\sin^2\phi
\end{equation}
The first term on the right hand side contributes to a term of the form
\begin{equation}
 \biggl( \frac{\omega_1}{\omega_0}\biggr)\theta_0 \phi\sin\phi\sim t\sin t
\end{equation}
which is secular, i.e. increases without bound. As we are looking for periodic solutions we must then have
\begin{equation}
 \omega_1= 0
\end{equation}
So, the solution to (\ref{eq:psiequation}) that satisfies the initial conditions
$\psi_1=0$ and $\psi_0'=0$ at $\phi = \omega t = 0$ is
\begin{equation}
 \psi_1 = \frac{1}{6}\theta_0^2 (3-4\cos\phi+\cos 2\phi )
\end{equation}
where we have used the identity $\sin ^2\phi = (1/2)(1-\cos 2\phi)$.
\noindent
From this result it follows that at the end of the first half-cycle ($\phi = \pi$) the amplitude will be
\begin{equation}
\label{eq:solutionquadratic}
 \theta_1 = -\theta_0\biggr( 1-\frac{4}{3}\epsilon\theta_0\biggr)
\end{equation}
This method can be applied successively to find the amplitudes of the next half-cycles. From this result it follows that the
difference between the first two successive amplitudes will be
\begin{equation}
\label{eq:diderotcompare}
 \theta_0-|\theta_1| =  \frac{4}{3}\epsilon(\theta_0)^2
\end{equation} 
If we translate Eq. (\ref{eq:diderotcompare}) into the language of arcs traversed by the bob, it is telling us that the difference
in arc is proportional to the square of the arc traversed by the bob during its descent. Apart from the prefactor of $4/3$, this
is the conclusion Diderot arrived at in his memoir.

\section{Conclusions}
\label{Conclusions}
In 1748 Diderot published a series of memoirs on different subjects of Mathematics. In the fifth memoir, he studied the effect
of air resistance on the movement of the pendulum when this resistance is proportional to the square of the bob's velocity.
Since Diderot quotes a passage of Newton's {\it Principia} where this problem is discussed considering a resistance
linear in the velocity, it has been argued in the past that the sole purpose of Diderot was to correct an assumption that Newton
made and Diderot thought incorrect. In the present article it has been argued that Diderot's
memoir may have served a different purpose: a careful analysis of his methods shows that Diderot wrote his memoir as a
detailed guide of how to get, in a mathematically rigorous way and using differential
calculus, the results that Newton presented without further justification. 

In order to do this he translated Newton's arguments into mathematical form and put them into a coherent mathematical framework.
He obtained an integral equation which he then solved by means of an approximation. By assuming the drag to be quadratic, he obtains a difference in amplitude
between swings quadratic in the displacement, a result which is confirmed by a Lindstedt-Poincar\'e analysis of the same problem.
By considering a drag linear in the velocity, he shows that Newton's results can be obtained in the limit of weak drag while giving
Newton's problem an exact solution.

The question of whether the drag should be linear or quadratic has been discussed in detail. If one considers the problem from a modern
perspective, the Reynolds number associated with a spherical bob of the size Newton used imply that drag should be quadratic, thus
confirming Diderot's assumption. However, from a practical point of view,
since the velocity varies during swings and one is usually interested in small amplitude oscillations, the difference in results obtained
in either case is beyond the precision that Newton had at his disposal and would not have been detected by Diderot in case he had conducted
himself the experiments.

Diderot handled a full problem for which even the simplified version (small-amplitude approximation with quadratic drag) had to wait
100 years to be appropriately handled. We may agree with Coolidge when he says that~\cite{coolidge2}
\begin{quote}
{\it '... Diderot had hold of a problem that was too much for him.'}
\end{quote}
However, as Eq. (\ref{eq:diderotcompare}) shows, this should not diminish his merit: he obtained the 
same functional dependence on the retardation as one would get using the modern perturbative approach by means of a first-order approximation
to solve an integral equation.

\section{Acknowledgements}

I would like to thank D. Hoffmann, J. Renn and C. Lehner from the Max Planck Institute for the
History of Science, Berlin, for their hospitality. I would also like to thank H. Hinrichsen from the University of W\"urzburg, D.
Wolf from the University of Duisburg-Essen, R. Kenna from Coventry University and P. Mac Carron from Oxford University for their interest and discussions.
N. Maillard native expertise in French is gratefully acknowledged.
Financial support from the Alexander-von-Humboldt Foundation, Germany and Project 295302 of the Marie-Curie-FP7-IRSES of the European
Community is gratefully acknowledged.

%
% References 
%
%


\begin{thebibliography}{99}
% 
%
\bibitem{diderot1} D Diderot, {\it M\'emoires sur diff\'erens sujets de Math\'ematiques}, Paris, 1748.
%
\bibitem{gay} P Gay, {\it The Enlightenment: An Interpretation. The Rise of Modern Paganism}, Vintage Books, New York, 1966, p. 14.
%%
\bibitem{kk} L G Krakeur, R L Krueger, {\it The Mathematical Writings of Diderot}, Isis 33/2 (1941), pp. 219 -- 232.
%%
\bibitem{coolidge} J L Coolidge, {\it The Mathematics of Great Amateurs}, 2nd edition, Clarendon Press, Oxford, 1990, pp. 178 -- 185.
%%
\bibitem{ballstadt} K Ballstadt, {\it Diderot: Natural Philosopher}, SVEC 2008:09, Oxford, Voltaire Foundation, 2008.
%
\bibitem{wilson} A M Wilson, {\it Diderot}, Oxford University Press, Oxford, 1972, pp. 18.
%
\bibitem{diderothelvetius} D Diderot, {\it Refutation suivie de l'ouvrage d'Helvetius intitule l'homme}, \'Editions Gallimard, Paris, 2010, p. 580.
%
\bibitem{wilson3} A M Wilson, op. cit., p. 22.
%
\bibitem{ballstadt2} K Ballstadt, {\it op. cit.} p. 9.
%
\bibitem{wilson2} A M Wilson, op. cit., p. 89.
%
\bibitem{huygenswork} C Huygens, {\it Horologium Oscillatorium: sive de motu pendulorum ad horologia aptato demonstrationes geometricae}, Culture et Civilisation,
Bruxelles, 1966.
%
\bibitem{jacsu} T Le Seur and F Jacquier, {\it Isaaci Newtoni Philosophiae Naturalis Principia Mathematica, Perpetuis Commentariis Illustrata},
Vol. I, Glasgow, 1822, pp. 37-39.
%
\bibitem{dahmen2} S R Dahmen,  {\it The Mathematics and Physics of
    Diderot II. On Involutes }, in preparation.
%
\bibitem{diderot3} D Diderot, {\it Oeuvres compl\`etes : \'edition critique et annot\'ee. V. 2: Philosophie et math\'ematique: Id\'ees I}. R. Niklaus {\it et. al.} (eds),
Herrmann, Paris, 1975.
%
\bibitem{wilson4} A M Wilson, op. cit., p. 49.
%
\bibitem{principia} I Newton, {\it Mathematical Principles of Natural Philosophy}, transl. Andrew Motte and F. Cajori. University of California
Press, Los Angeles, 1934. Reprinted as vol. 34 of Great Books of the Western
World Series, Encyclopedia Britannica, Chicago, 1952, pp. 20-21.
%
\bibitem{mayer} J Mayer in ~\cite{diderot3}, p. 321.
%
\bibitem{gauld1} C F Gauld, {\it Newton’s Use of the Pendulum to Investigate Fluid
Resistance: A Case Study and some Implications for Teaching About the Nature of Science}, Science and Education 18 (2009), 383–400.
%%
\bibitem{gauld2} C F Gauld, {\it Newton’s Investigation of the Resistance to Moving
Bodies in Continuous Fluids and the Nature
 of `Frontier Science'}, Science and Education 19 (2010), 939–961.
%
\bibitem{wilson5} A M Wilson, op. cit., p. 69.
%
\bibitem{principia3} I Newton, {\it op. cit.} p. 165.
%
\bibitem{good} R H Good, {\it Elliptic Integrals, the forgotten functions}, Eur. J. Phys {\bf 22} (2001), 119 - 126.
%
\bibitem{fluegge} S Fl\"ugge, {\it Lehrbuch der Theoretischen Physik Band II}, Springer Verlag, Heidelberg, 1967.
%\bibitem{ct} C Cohen-Tannoudji, B Liu, F Laloe, {\it Quantum Mechanics}, Wiley-VCH, Weinheim, 1992.
%%
\bibitem{linz} Linz, S.J. 1995. {\it A simple nonlinear system: the pendulum with quadratic friction}, Eur. J. Phys. {\bf 16}, pp. 67 -- 72.
%
\bibitem{no} R A Nelson, M G Olsson, {\it The pendulum -- rich physics from a simple system}, Am. J. Phys. {\bf 54}/2 (1986), pp. 112 -- 121.
%
\bibitem{cve} Cveti\'canin, L. 2009. {\it Oscillator with strong quadratic damping force}, Publications de l'Institute Math\'ematique, Nouvelle
s\'erie, tome 85(99), 119 -- 130.
%
\bibitem{stokes} G G Stokes, {\it On the Effect of the Internal Friction of Fluids on the Motion of Pendulums},
Transactions of the Cambridge Philosophical Society 9  (1851), pp. 8--106.
%%
\bibitem{reynoldscite} O Reynolds, {\it An experimental investigation of the circumstances which determine whether the motion
of water shall be direct or sinuous, and of the law of resistance in parallel channels}, Philosophical Transactions of the Royal Society 174 (0) (1883), pp. 935–-982.
%%
\bibitem{white} F M White, {\it Viscuous Fluid Flow}, McGraw-Hill, New York, 1974, p. 209.
%%
\bibitem{lindstedt} A Lindstedt,{\it Beitrag zur Integration der Differentialgleichungen der St\"orungstheorie}, 
Abh. K. Akad. Wiss. St. Petersburg 4, 31 (1882).
%%
\bibitem{poincare} H Poincare, {\it Les M\'ethodes Nouvelles de la M\'ecanique C\'el\`este II}, Gauthier-Villars Imprimieus Librarie, Paris, 1893 paragraph 123 - 128 (1893).
%%
\bibitem{coolidge2} J L Coolidge, {\it op. cit.} p. 184.
%\bibitem{dalembert} D'Alembert, {\it Discours Preliminaire}
\bibitem{bessel} F W Bessel, {\it Untersuchungen \"uber die L\"ange des einfachen Secundenpendels}, Druckerei der K\"oniglichen Akademie der Wissenschaften,
Berlin, 1828. Reprint Verlag von
Wilhelm Engelmann, Leipzig, 1889.
%\bibitem{webster} Some definitions of 'native' taken from the Webster Dictionary: natural, unnafected, indigenous, innate.
%\bibitem{diderot2} Prospectus
%\bibitem{clive} Emsley C, Hitchcock T and Shoemaker R, "London History - Currency, Coinage and the Cost of Living",
%Old Bailey Proceedings Online, http:www.oldbaileyonline.org. Retrieve April 30 2014.
%\bibitem{kafker}
%\bibitem{darnton} Darnton R, {\it The Great Cat Massacre and Other Episodes in French Cultural History}, Basic Books, New York, 1999.
\end{thebibliography}
\end{document}